\documentclass{article}
\usepackage{moreverb}
\usepackage[colorlinks,bookmarksopen,bookmarksnumbered,citecolor=red,urlcolor=red]{hyperref}
\usepackage{amsmath, amsfonts, mathtools}
\usepackage{authblk}
\usepackage{booktabs}
\usepackage{setspace}
\usepackage[a4paper, total={6in, 8in}]{geometry}

\begin{document}

\title{One-Sample Log-Rank Tests with Consideration of Reference Curve Sampling Variability}

\author[1,2]{Moritz Fabian Danzer}
\author[1,2]{Jannik Feld}
\author[1]{Andreas Faldum}
\author[1]{Rene Schmidt}
\affil[1]{Institute of Biostatistics and Clinical Research, University of M\"unster, 48149 M\"unster, Germany}
\affil[2]{Authors contributed equally}

\maketitle

\begin{abstract}
The one--sample log--rank test is the method of choice for single--arm Phase II trials with time--to--event endpoint. It allows to compare the survival of the patients to a  reference survival curve that typically represents the expected survival under standard of care. The classical one--sample log--rank test, however, assumes that the reference survival curve is deterministic. This ignores that the reference curve is commonly estimated from historic data and thus prone to statistical error. Ignoring sampling variability of the reference curve results in type I error rate inflation.
For that reason, a new one--sample log--rank test is proposed that explicitly accounts for the statistical error made in the process of estimating the reference survival curve.
The test statistic and its distributional properties are derived using martingale techniques in the large sample limit. In particular, a sample size formula is provided. Small sample properties regarding type I and type II error rate control are studied by simulation. A case study is conducted to study the influence of several design parameters of a single-armed trial on the inflation of the type I error rate when reference curve sampling variability is ignored.
\end{abstract}

\section{Introduction}\label{Sec_01}
The one--sample log--rank test is the method of choice for single--arm Phase II trials with time--to--event endpoint. It allows to compare the survival of the patients to a prefixed reference survival curve that typically represents the expected survival under standard of care. First proposed by \cite{Breslow:1975}, its practical implementation including sample size calculation has been described by \cite{Finkelstein:2003}.
The one--sample log--rank test is often criticized in different directions. First, it has been reported repeatedly in the literature that the classical one--sample log--rank statistic tends to be conservative (see \cite{Wu:2014, Schmidt:2015}).
One reason for the test's inaccuracy is the dependence between the estimators of mean and variance of the original one--sample log--rank statistic when sample size is small.
Several attempts have been made in the literature to correct for this (see\cite{Wu:2014, Schmidt:2015, Sun:2011, Wu:2015, Wu:2016, Kerschke:2020}).
Amongst those, the proposal made by \cite{Wu:2015} is presently implemented in the commercial software PASS \cite{PASS:2018} for sample size calculation for the one--sample log--rank test.
%Recently, \cite{Danzer:2021a} proposed a general framework that enables further improvement of the proposal made by \cite{Wu:2015}. 
%
Another more conceptual point of criticism against the one--sample log--rank test relates to the process of selecting the reference survival curve. It is common practice to choose the reference survival curve in the light of historic data on standard treatment. This implies that choice of the reference survival curve itself is thus prone to statistical error which, however, is ignored in the classical one--sample log--rank statistic. As lined out in \cite{Korn:2006}, this is as general problem in clinical trials with historical controls. Accordingly, common one--sample log--rank tests rather assume that the reference survival curve is a priori known and deterministic as in \cite{Finkelstein:2003, Wu:2014, Schmidt:2015, Sun:2011, Wu:2015, Wu:2016, Kerschke:2020}. This ignores that the reference curve resulted from an estimation process and complicates interpretation of the test results. Moreover, historic data often suffer from not reflecting recent advances in diagnostics and/or concomitant therapy for standard of care.

To overcome these interpretative limitations we propose a new one--sample log--rank test that explicitly accounts for statistical error made in the process of estimating and fixing the reference survival curve. Principally, the new test applies to both historic and prospective comparisons of a new treatment to a standard in the framework of Phase II survival trials. In the latter case, the new test may also be interpreted as a two--sample test for survival distributions.

The paper is organized as follows. After settling notation and the testing problem, we describe the test statistic and it distributional properties. Additionally, we provide sample size calculation methods. Calculation of rejection regions and sample size are based on the approximate distribution of the new test statistic in the large sample limit. Therefore small sample properties of the new test regarding type I and type II error rate control are studied by simulation, and compared to the classical one--sample log--rank test / two--sample log--rank test. These simulations and a case study shed light on the inflation of the type I error rate that results from ignoring the sampling variability of the reference curve in the planning phase of a new single-armed trial. We conclude with a discussion of future research. Mathematical proofs are shifted to Appendix A.

\section{General Aspects}\label{Sec_02}

\subsection{Notation}\label{Sec_02_01}

We consider a survival trial with survival data from two treatment groups A (control intervention, prospectively collected or historic data) and B (experimental intervention, prospectively collected data). Let ${\cal{N}}_x$ denote the set of patients from group $x=A,B$, $n_x\coloneqq|{\cal{N}}_x|$ the number of such patients, and $n\coloneqq n_A+n_B$ the total number of patients. In particular, we denote by $\pi=n_B/n_A$ the treatment group allocation ratio. We denote by $T_{x,i}$ or $C_{x,i}$ the time from entry to event or censoring for patient $i$ from group $x=A,B$, respectively. Let $X_{x,i}\coloneqq T_{x,i} \wedge C_{x,i}$ denote the minimum of both. As usual, we assume that the $T_{x,i}$ and $C_{x,i}$ are mutually independent (non--informative censoring). For each $s \geq 0$, we denote by ${\cal{F}}_s$ the $\sigma$--algebra of information available by study time $s$:

\begin{align}\label{Sec_02_01_eq01}
{\cal{F}}_s\coloneqq
\sigma \left( I{(T_{x,i} \leq s) }, T_{x,i} \cdot I{(T_{x,i} \leq s)}, I{(C_{x,i} \leq s)}, C_{x,i} \cdot I{(C_{x,i} \leq s)}; i=1,\ldots,n_x, x=A,B \right).
\end{align}

Based on the observed data, we calculate the \emph{number of events} from treatment group $x=A,B$ up to study time $s \geq 0$ as

\begin{align}\label{Sec_02_01_eq02}
\begin{split}
N_{x}(s)\coloneqq \sum_{i \in {\cal{N}}_{x}} N_{x,i}(s), \quad N_{i}(s)\coloneqq I( T_{x,i} \leq s, T_{x,i} \leq C_{x,i}), 
\end{split}
\end{align}

and the \emph{number at risk} $Y_{x}(s)\coloneqq \sum_{i \in {\cal{N}}_{x}} I( T_{x,i} \wedge C_{x,i} \geq s  )$ by study time $s \geq 0$ in treatment group $x=A,B$. 
%\begin{align*}%\label{Sec_02_01_eq03}
%\begin{split}
%\end{split}
%\end{align*}
Let $J_{x}(s)\coloneqq I( Y_{x}(s)> 0)$ indicate whether there are still patients at risk in treatment group $x$ by study time $s$. As usual, we let $\lambda_x(s)\coloneqq \lim_{\Delta \to 0} P(s \leq T_{x,i} < s + \Delta| T_{x,i} \geq s)/\Delta$ denote the hazard of a patient $i$ from treatment group $x=A,B$. We denote by $\Lambda_x(s)\coloneqq \int_0^s \lambda_x(u)du$ the corresponding cumulative hazard function for treatment group $x=A,B$, respectively. 
Finally, we denote by $f_{T_x}$, $F_{T_x}$, $S_{T_x}$ (or $f_{C_x}$, $F_{C_x}$, $S_{C_x}$) the density, distribution function and survival function of the time to event $T_{x,i}$ (or time to censoring $C_{x,i}$) in treatment group $x=A,B$. Notice that $\lambda_x$, $\Lambda_x$, $f_{T_x}$, $F_{T_x}$, $S_{T_x}$ and $f_{C_x}$, $F_{C_x}$, $S_{C_x}$ are assumed to coincide for all patients from the same treatment group $x=A,B$. 

We will also need the Nelsen--Aalen estimator 

\begin{align}\label{Sec_02_01_eq04}
\begin{split}
\widehat{\Lambda}_x(s)
\coloneqq \int_0^s \frac{J_x(u)}{Y_x(u)} dN_x(u)
\equiv \ \sum_{\substack{ i \in {\cal{N}}_{x}, \\ N_{x,i} (s) = 1  }}   \frac{J_x(T_{x,i})}{Y_x(T_{x,i})}
\end{split}
\end{align}

of the cumulative hazard function $\Lambda_x(s)$ for group $x=A,B$, and the corresponding estimator of the variance function

\begin{align}\label{Sec_02_01_eq05}
\begin{split}
\widehat{\sigma}_x(s)
\coloneqq n_x \cdot \int_0^s \frac{J_x(u)}{Y_x^2(u)} dN_x(u) \
\equiv \ n_x \cdot \sum_{\substack{ i \in {\cal{N}}_{x}, \\ N_{x,i} (s) = 1  }}   \frac{J_x(T_{x,i})}{Y_x^2(T_{x,i})}.
\end{split}
\end{align}

We consider $N_{x}$, $Y_{x}$, $J_{x}$, $\widehat{\Lambda}_x$ and $\widehat{\sigma}_x$ as stochastic processes in study time $s \geq 0$, adapted to the filtration $({\cal{F}}_s)_{s \geq 0}$. 
Notice that we define $0/0\coloneqq0$ whenever formal division of zero by zero occurs in a mathematical expression.

%======================================================================================================================
\subsection{The Testing Problem}\label{Sec_02_02}
%======================================================================================================================

We consider testing the null hypothesis that the survival function of patients from the experimental treatment group $B$ coincides with the reference curve that is given by the \emph{true} survival function under standard of care, i.e.

\begin{equation*}
H_0: \Lambda_B(s) = \Lambda_A(s)  \text{ for all } s \in [0,s_{max}] 
\end{equation*}

for some maximum observation time $s_{max}>0$.\\
Notice that $H_0$ deviates from the null hypothesis of the classical one-sample log-rank tests (see \cite{Breslow:1975} or \cite{Finkelstein:2003}) which assumes a known reference survival curve. Nevertheless, $H_0$ typically is the null hypothesis of actual interest also in a one-sample setting.

%======================================================================================================================
\section{The Testing Procedure}\label{Sec_03}
%======================================================================================================================

%======================================================================================================================
\subsection{Motivation}\label{Sec_03_01}
%======================================================================================================================
Starting point is the stochastic process $M_0(s)\coloneqq n_B^{-1/2}[N_B(s) - \sum_{i \in {\cal{N}}_{B}} \Lambda_A(s \wedge X_{B,i})]$. When $H_0$ holds true, $M_0$ is (known to be) a mean--zero ${\cal{F}}_s$--martingale. $M_0$ depends on data from the experimental treatment arm $B$, only, and is commonly used as a basis to construct one--sample log--rank tests (see \cite{Aalen:2008}). Notice, however, the difficulty that $M_0$ depends on the true unknown cumulative hazard function $\Lambda_A$ under standard of care. In the context of the classical one--sample log--rank test it is common practice to estimate $\Lambda_A$ from historic data, and to identify the obtained estimate $\widetilde{\Lambda}_A$ with $\Lambda_A$, while treating $\widetilde{\Lambda}_A$ as a deterministic function. I.e., the classical one--sample log--rank test effectively assesses the null hypothesis $\widetilde{H}_0:\Lambda_B = \widetilde{\Lambda}_A$ using the test statistic $\widetilde{M}_0(s)\coloneqq n_B^{-1/2}[N_B(s) - \sum_{i \in {\cal{N}}_{B}} \widetilde{\Lambda}_A(s \wedge X_{B,i})]$ while pretending that $\widetilde{\Lambda}_A$ is an a priori known deterministic reference function representing the expected survival under standard of care. This, however, may detract from the actually interesting null hypothesis $H_0: \Lambda_B = \Lambda_A$ when random deviation of $\widetilde{\Lambda}_A$ from $\Lambda_A$ is large.
To avoid those interpretive difficulties, we here propose to incorporate the process of reference curve estimation into the one--sample log--rank statistic: Replacing $\Lambda_A$  with its Nelsen--Aalen estimate $\widehat{\Lambda}_A$ (see \cite{Nelson:1969, Nelson:1972,Aalen:1978}) in the definition of $M_0$ while treating $\widehat{\Lambda}_A$ as random, we obtain a new stochastic process $\widehat{M}_0(s)\coloneqq n_B^{-1/2}[N_B(s) - \sum_{i \in {\cal{N}}_{B}} \widehat{\Lambda}_A(s \wedge X_{B,i})]$ that (i) can be calculated from the data, and (ii) may be used as test statistic for the original null hypothesis $H_0$ as we will see below. Notice that replacing $\Lambda_A$ with $\widehat{\Lambda}_A$ in $M_0$ increases the variance of the stochastic process since $\widehat{\Lambda}_A$ contributes additional variability. Deriving the correct rejection regions thus requires separate consideration which is not covered by the underlying one--sample test methodology. The resulting significance test may also be interpreted as a two--sample survival test, as the reference curve coincides with the true survival function under standard of care. Our proceeding defines a general strategy to lift existing methodology for one--sample survival tests to a multi--sample setting for a variety of different design settings as will be further discussed.

%======================================================================================================================
\subsection{Test Statistic and Significance Test}\label{Sec_03_02}
%======================================================================================================================
Consider the ${\cal{F}}_s$--adapted stochastic processes

\begin{align}\label{Sec_03_02_eq01}
\begin{split}
\widehat{M}_0(s) &\coloneqq n_B^{-1/2}\left[N_B(s) - \sum_{i \in {\cal{N}}_{B}} \widehat{\Lambda}_A(s \wedge X_{B,i})\right] \\
\widehat{\Sigma}^2(s) &\coloneqq n_B^{-1} N_B(s) + n_B^{-1} n_A^{-1} \sum_{i,j \in {\cal{N}}_{B}} \widehat{\sigma}_A(s \wedge X_{B,i} \wedge X_{B,j})
\end{split}
\end{align}
with $N_B$, $\widehat{\Lambda}_A$ and $\widehat{\sigma}_A$ acc. to (\ref{Sec_02_01_eq02}), (\ref{Sec_02_01_eq04}) and (\ref{Sec_02_01_eq05}). Assume that the null hypothesis $H_0$ holds true. Then by Theorem 1 (see Appendix A) the following applies: (i) $\widehat{M}_0$ is a mean--zero ${\cal{F}}_s$--martingale with asymptotically independent increments, i.e. for any $0<s_1<s_2 < s_{max}$, $\widehat{M}_0(s_1)$ and $\widehat{M}_0(s_2)-\widehat{M}_0(s_1)$ are approximately independent when sample size $n$ is sufficiently large, and (ii) for each fixed $s>0$ we have $\widehat{M}_0(s) \stackrel{d}{\to} {\cal{N}}(0,\Sigma^2(s))$ in distribution as $n\to\infty$, where $\Sigma(s)\coloneqq\text{plim}_{n \to \infty} \widehat{\Sigma}(s) = \lim_{n\to\infty} E[\widehat{\Sigma}(s)]$ (see Appendix A, Lemma 1).
In particular, the random variable

\begin{align}\label{Sec_03_02_eq02}
\begin{split}
Z\coloneqq \frac{\widehat{M}_0(\infty)}{\widehat{\Sigma}(\infty)} \stackrel{H_0}{\sim} {\cal{N}}(0,1)
\end{split}
\end{align}
is approximately standard normally distributed under the null hypothesis $H_0$. Notice that the parameters $n_A$ and $n_B$ cancel out in the definition of $Z$, so that $Z$ can be calculated from the observed data. Thus an approximate level $\alpha$ test of $H_0$ is defined by rejecting $H_0$ whenever $|Z| \geq \Phi^{-1}(1-\alpha/2)$, where $\alpha$ is the desired two--sided significance level and $\Phi$ is the standard normal distribution function.

To enable easy application of the proposed significance test in clinical practice, we provide R code (see \cite{R}) that calculates the value of the test statistic $Z$ and the corresponding two--sided $p$--value for given input data set. The R code as well as instructions how to prepare the input data are given in \nameref{S1_File}.  

%======================================================================================================================
\section{Sample Size Calculation}\label{Sec_04}
%======================================================================================================================

Sample size is calculated under the proportional hazards planning alternative $K_0: \Lambda_B(s) = \omega_0 \cdot \Lambda_A(s)$ for some prefixed hazard ratio $0<\omega_0<1$. By Theorem 2 (see Appendix A), the test statistic $Z$ from (\ref{Sec_03_02_eq02}) is approximately normally distributed under planning alternative hypothesis $K_0$ with unit variance and mean $\log(\omega_0) \cdot \mu \cdot \sigma^{-1}$ where 

\begin{align}\label{Sec_04_eq01}
\begin{split}
\mu &= n^{1/2} \sqrt{\frac{\pi}{1+\pi}} \int_0^{\infty} F_{T_A}(u) f_{C_A}(u) du %= n^{-1/2} \sqrt{\frac{1+\pi}{\pi}} E[N_{A}(\infty)]
\\
\sigma^2 &= \int_0^{\infty} F_{T_A}(u) f_{C_A}(u) du + 2 \pi \cdot \int_0^{\infty} \sigma_A(u) [f_{T_A}(u)S_{C_A}(u)+S_{T_A}(u)f_{C_A}(u)]S_{T_A}(u)S_{C_A}(u) du
\end{split}
\end{align}
with ${\sigma}_A(s) \coloneqq \int_0^s \frac{\lambda_A(u)}{S_{T_A}(u) \cdot S_{C_A}(u)}  du$ and $\pi=n_B/n_A$ denoting the treatment arm allocation ratio. Large negative values of the test statistic $Z$ support validity of the planning alternative $K_0$. The power $1-\beta\coloneqq P_{K_0}(Z \leq \Phi^{-1}(\alpha/2))$ of the trial is thus approximately given by

\begin{align}\label{Sec_04_eq02}
\begin{split}
1- \beta = \Phi \left( \Phi^{-1}(\alpha/2) - \log(\omega_0) \cdot \mu \cdot \sigma^{-1} \right).
\end{split}
\end{align}
In practice, the following assumptions on accrual and censoring are commonly made when calculating the required sample size of a survival trial:
\begin{itemize}
	\item Patients enter the trial uniformly between year $0$ and year $a$ with prefixed constant accrual rate $r$, say, and are then followed--up for further $f \geq s_{max}$ years until the time of final analysis in year $a+f$ year.
	\item No loss to follow--up, i.e. $C_{x,i}\sim {\cal{U}}(f,a+f)$ is uniformly distributed on $[f,a+f]$.
\end{itemize}
These assumption amount to
\begin{align}\label{Sec_04_eq03}
\begin{split}
n &= r \cdot a \\
f_{C_x}(s) &= a^{-1} \cdot I\left( s \in [f,a+f] \right),  \qquad \text{for } s \geq  0, \\
S_{C_x}(s) &= \min\left(1, \frac{a+f-s}{a}\right) \cdot I\left( s \leq a+f \right), \qquad \text{for } s \geq  0, \\
\sigma_A(s) &= \int_0^{s} \frac{\lambda_A(u)}{S_{T_A}(u)} \max\left(1, \frac{a}{a+f-u}\right) du  \qquad \text{for } 0 \leq s < a+f.
\end{split}
\end{align}
thus further simplifying above expressions for $\mu$ and $\sigma$.
For prefixed two--sided significance level $\alpha$, hazard ratio $0<\omega_0<1$, treatment group allocation ratio $\pi=n_B/n_A$, overall accrual rate $r$, length of the follow--up period $f$, and control arm cumulative hazard function $\Lambda_A(s)$, it remains to choose the only remaining free parameter $a$ in (\ref{Sec_04_eq02}) such that a desired power $1-\beta$ is achieved. With the parameter $a$ calculated this way, the required number of patients $n$ to achieve a power of $1-\beta$ under planning alternative $K_0$ is $n=r \cdot a$.

In \nameref{S1_File} we provide R code (see \cite{R}) that calculates the required number of patients $n$ for settings when the survival times in the reference group $A$ are Weibull distributed $\Lambda_A(s)\coloneqq - \log(S_1) \cdot t^{\kappa}$ with prefixed shape parameter $\kappa$ and prefixed 1-year survival rate $S_A(1)=S_1$. 

\section{Simulation Study I: Comparison with the Classical One--Sample Log--Rank Test}\label{Sec_06}
%======================================================================================================================

%======================================================================================================================
\subsection{Design}\label{Sec_06_01}
%======================================================================================================================

In the application of the classical one--sample log--rank test from \cite{Breslow:1975,Finkelstein:2003} it is common practice to estimate the standard arm hazard function $\Lambda_A$ from historic data, and to choose the obtained estimate $\widetilde{\Lambda}_A$ as the reference curve, while treating $\widetilde{\Lambda}_A$ as deterministic. This may lead to  type I error rate inflation when the underlying null hypothesis to be tested is $H_0: \Lambda_B = \Lambda_A$, because the random deviation of $\widetilde{\Lambda}_A$ from $\Lambda_A$ is neglected and the variance of the involved test statistics is thus underestimated. The objective of this simulation study I is to quantify the amount of type I error rate inflation in settings of clinical relevance: We study and compare the empirical type I error rates when (i) the classical one--sample log--rank test (without correction for sampling variability of the reference curve) and (ii) the new  one--sample log--rank test (with correction for sampling variability of the reference curve) is used to test null hypotheses $H_0: \Lambda_B = \Lambda_A$.

In our simulations, patients were assumed to enter the trial uniformly between year $0$ and year $a$ with overall accrual rate of $r=100$ per year.
Accordingly the calendar times of entry were generated according to a uniform distribution $Y_{x,i} \sim {\cal{U}}(0,a)$ on $[0,a]$. 
%Patients were allocated equally to both treatment arms $A$ and $B$ (allocation ratio 1:1), corresponding to an annual accrual rate of $50$ patients per group.
After the end of the accrual period, patients were assumed to be followed up for further $f=3$ years, while assuming no loss to follow--up. 
Accordingly, we set $C_{x,i}\coloneqq a+f-Y_{x,i}\sim {\cal{U}}(f,a+f)$. 
Survival times $T_{A,i}$ in the control intervention group $A$ were generated according to a Weibull distribution $\Lambda_A(s)\coloneqq - \log(S_1) \cdot t^{\kappa}$ with prefixed shape parameter $\kappa$ and 1-year survival rate $S_{T_A}(1)=S_1=0.5$. Survival times $T_{B,i}$ in the experimental intervention group $B$ were generated acc. to a Weibull distribution with $\Lambda_B(s)\coloneqq \omega \cdot \Lambda_A(s)$, where $\omega$ is the true hazard ratio. 

To perform the classical one--sample log--rank test, the standard arm data was used to calculate the Nelsen--Aalen estimate $\widehat{\Lambda}_A$ of $\Lambda_A$. The obtained estimate $\widehat{\Lambda}_A$ was then treated as a deterministic function and used as (prefixed, deterministic) reference cumulative hazard function in the classical one--sample log--rank statistic (see \cite{Breslow:1975,Finkelstein:2003}). In contrast, the new test was performed according to our previously shown results.

To study the impact of sample size and allocation ratio on the amount of type I error rate inflation, the total sample size $n = r \cdot a$ of the virtual data sets was chosen as $n= 100, 500, 1000$. For each of these total sample sizes we considered allocation ratios $\pi \in \{2, 1, 1/2, 1/4, 1/8, 1/16\}$. Scenarios with $\pi \leq 1/2$ are more likely to reflect common practice as the size of the experimental cohort is typically smaller than the size of the historical control cohort. To study the impact of different shapes of the survival distribution, we considered different values for the Weibull shape parameter from the interval $[0.1,5]$.

For each parameter constellation, we generated 10000 samples of size $n$ to which we applied both the new test as well as the classical one--sample log--rank test. The desired two--sided significance level was $5\%$. Results are shown in Table \ref{tab:AlphaInflation} and discussed below.

\begin{table}[!ht]
	\small
		\centering
		\caption{\bf Empirical type I error rates under consideration of sampling variability}
		\begin{tabular}{cc||cc|cc|cc|cc|cc|cc}
			\hline
			&  & 
			\multicolumn{2}{c|}{$\pi = 2$ } &
			\multicolumn{2}{c|}{$\pi = 1$ } &
			\multicolumn{2}{c|}{$\pi = 1/2$ } &
			\multicolumn{2}{c|}{$\pi = 1/4$ } &
			\multicolumn{2}{c|}{$\pi = 1/8$ } &
			\multicolumn{2}{c}{$\pi = 1/16$ } \\[0.14cm]
			$\kappa$ & $n$ & $\alpha_{new}$ & $\alpha_{LR}$& $\alpha_{new}$ & $\alpha_{LR}$& $\alpha_{new}$ & $\alpha_{LR}$& $\alpha_{new}$ & $\alpha_{LR}$& $\alpha_{new}$ & $\alpha_{LR}$& $\alpha_{new}$ & $\alpha_{LR}$ \\[0.14cm] 
			\hline
			0.1 &  100 & 0.041 & 0.262 & 0.044 & 0.170 & 0.047 & 0.113 & 0.055 & 0.084 & 0.064 & 0.075 & 0.107 & 0.071 \\ 
			0.1   &  500 & 0.044 & 0.249 & 0.047 & 0.163 & 0.052 & 0.112 & 0.053 & 0.084 & 0.050 & 0.062 & 0.056 & 0.062 \\ 
			0.1   & 1000 & 0.047 & 0.257 & 0.046 & 0.160 & 0.049 & 0.106 & 0.051 & 0.079 & 0.054 & 0.068 & 0.051 & 0.059 \\[0.14cm] 
			0.25 &  100 & 0.043 & 0.267 & 0.045 & 0.172 & 0.047 & 0.114 & 0.053 & 0.086 & 0.062 & 0.075 & 0.086 & 0.070 \\ 
			0.25   &  500 & 0.047 & 0.252 & 0.049 & 0.162 & 0.049 & 0.111 & 0.052 & 0.082 & 0.050 & 0.065 & 0.057 & 0.065 \\ 
			0.25   & 1000 & 0.048 & 0.259 & 0.048 & 0.161 & 0.049 & 0.106 & 0.052 & 0.081 & 0.051 & 0.066 & 0.053 & 0.058 \\[0.14cm] 
			0.5 &  100 & 0.044 & 0.278 & 0.048 & 0.174 & 0.049 & 0.118 & 0.052 & 0.090 & 0.058 & 0.077 & 0.069 & 0.078 \\ 
			0.5  &  500 & 0.049 & 0.260 & 0.050 & 0.168 & 0.051 & 0.112 & 0.052 & 0.083 & 0.056 & 0.068 & 0.056 & 0.067 \\ 
			0.5   & 1000 & 0.051 & 0.263 & 0.052 & 0.164 & 0.052 & 0.112 & 0.055 & 0.083 & 0.053 & 0.068 & 0.055 & 0.061 \\[0.14cm] 
			0.75 &  100 & 0.043 & 0.289 & 0.050 & 0.184 & 0.049 & 0.126 & 0.053 & 0.093 & 0.056 & 0.080 & 0.056 & 0.082 \\ 
			0.75   &  500 & 0.047 & 0.272 & 0.051 & 0.173 & 0.051 & 0.116 & 0.052 & 0.084 & 0.058 & 0.076 & 0.052 & 0.066 \\ 
			0.75  & 1000 & 0.053 & 0.269 & 0.053 & 0.171 & 0.052 & 0.114 & 0.054 & 0.084 & 0.050 & 0.067 & 0.051 & 0.062 \\[0.14cm] 
			1 &  100 & 0.036 & 0.300 & 0.051 & 0.194 & 0.049 & 0.130 & 0.052 & 0.097 & 0.052 & 0.082 & 0.052 & 0.085 \\ 
			1   &  500 & 0.047 & 0.273 & 0.050 & 0.176 & 0.052 & 0.116 & 0.052 & 0.085 & 0.053 & 0.073 & 0.052 & 0.069 \\ 
			1  & 1000 & 0.053 & 0.269 & 0.052 & 0.170 & 0.051 & 0.115 & 0.052 & 0.084 & 0.051 & 0.069 & 0.050 & 0.063 \\[0.14cm] 
			1.25 &  100 & 0.026 & 0.293 & 0.043 & 0.197 & 0.043 & 0.133 & 0.050 & 0.097 & 0.049 & 0.084 & 0.039 & 0.087 \\ 
			1.25  &  500 & 0.047 & 0.275 & 0.051 & 0.177 & 0.049 & 0.116 & 0.052 & 0.086 & 0.050 & 0.072 & 0.049 & 0.066 \\ 
			1.25   & 1000 & 0.051 & 0.270 & 0.051 & 0.171 & 0.052 & 0.114 & 0.053 & 0.085 & 0.049 & 0.068 & 0.050 & 0.063 \\[0.14cm] 
			1.5 &  100 & 0.023 & 0.270 & 0.036 & 0.185 & 0.037 & 0.128 & 0.043 & 0.097 & 0.037 & 0.082 & 0.026 & 0.085 \\ 
			1.5 &  500 & 0.046 & 0.276 & 0.052 & 0.177 & 0.049 & 0.116 & 0.051 & 0.086 & 0.050 & 0.072 & 0.048 & 0.066 \\ 
			1.5  & 1000 & 0.050 & 0.271 & 0.051 & 0.172 & 0.051 & 0.116 & 0.053 & 0.084 & 0.048 & 0.067 & 0.048 & 0.064 \\[0.14cm] 
			2 &  100 & 0.024 & 0.261 & 0.034 & 0.174 & 0.032 & 0.122 & 0.035 & 0.090 & 0.029 & 0.077 & 0.015 & 0.082 \\ 
			2  &  500 & 0.045 & 0.276 & 0.051 & 0.177 & 0.049 & 0.116 & 0.050 & 0.087 & 0.049 & 0.072 & 0.047 & 0.065 \\ 
			2  & 1000 & 0.049 & 0.271 & 0.052 & 0.172 & 0.050 & 0.116 & 0.053 & 0.084 & 0.048 & 0.067 & 0.048 & 0.064 \\[0.14cm] 
			5 &  100 & 0.024 & 0.261 & 0.034 & 0.173 & 0.032 & 0.122 & 0.035 & 0.090 & 0.029 & 0.077 & 0.014 & 0.082 \\ 
			5  &  500 & 0.045 & 0.275 & 0.051 & 0.177 & 0.049 & 0.116 & 0.050 & 0.087 & 0.049 & 0.071 & 0.047 & 0.065 \\ 
			5   & 1000 & 0.049 & 0.271 & 0.052 & 0.172 & 0.050 & 0.116 & 0.052 & 0.084 & 0.048 & 0.067 & 0.048 & 0.064 \\ 
			\hline
		\end{tabular}
		\begin{flushleft} Empirical type I error rates $\alpha_{new}$ and $\alpha_{LR}$ for testing $H_0:\Lambda_B =  \Lambda_A$ using the new test statistic $Z$ and the classical one--sample log--rank statistic, respectively, for Weibull distributed survival times with shape parameter $\kappa$ and 1--year survival rate $S_1=0.5$ in the control arm. Theoretical two--sided significance level: $5 \%$. Underlying total sample size of $n$ with allocation ratio $\pi$. 
		\end{flushleft}
		\label{tab:AlphaInflation}
\end{table}

%======================================================================================================================
\subsection{Results}\label{Sec_06_02}
%======================================================================================================================

The classical one--sample log--rank test does not account for sampling variability of the reference curve estimate. This leads to type I error rate inflation when the underlying null hypothesis to be tested is $H_0: \Lambda_B = \Lambda_A$. As expected, our simulations support that the amount of type I error rate inflation of the classical one--sample log--rank test is most pronounced when the allocation ratio $\pi$ is large. For any fixed allocation ratio, the inflation slightly decreases with increasing overall sample size but remains on a similar level. For ratios $\pi \geq 1$, the true type I error rate is more than three times higher than the desired one ($\sim 17\%$ instead of $5\%$ for $\pi=1$). For low allocation ratios as $1/8$ or $1/16$, the actual type I error still exceeds the nominal level, but to an extent that might be acceptable for a phase II trial ($\sim 6.5\%$ for $\pi = 1/16$ and $n = 1000$). With a view to the classical one-sample log-rank test, this supports that choice of the reference curve should be based on a historic control that is at least 10 times larger than the new experimental trial cohort. Reassuringly, the new test that explicitly accounts for reference curve variability realizes an empirical type I error rate close to the desired $5\%$ in almost all scenarios. Notice that the new test would hardly be applied in the scenario with $n=100$ and $\pi = 1/16$ as this implies a trial with $n_B=100/17 \approx 6$ only. So the entries for $n=100$ and $\pi = 1/16$ have to be interpreted with care, but are shown for reasons transparency and completeness.
The simulations thus support that neglecting the reference curve variability relevantly compromises type I error rate control when testing null hypotheses $H_0: \Lambda_B = \Lambda_A$. Notice that the classical one--sample log--rank test only realizes strict type I error rate control for testing the null hypothesis $\widetilde{H}_0:\Lambda_B = \widetilde{\Lambda}_A$ which, however, detracts from the null hypothesis $H_0: \Lambda_B = \Lambda_A$ when random deviation of $\widetilde{\Lambda}_A$ from $\Lambda_A$ is large.

%======================================================================================================================
\section{Simulation Study II: Comparison with the Two--Sample Log--Rank Test}\label{Sec_05}
%======================================================================================================================

%======================================================================================================================
\subsection{Design}\label{Sec_05_01}
%======================================================================================================================

We proposed a significance test for null hypothesis $H_0$ based on the approximate large sample distribution of the test statistic $Z$ introduced before. Despite of being derived as a one--sample log--rank test with consideration of reference curve variability, the new test may also be interpreted as a two--sample survival test. This simulation therefore aims to study performance of the new survival test for sample sizes of practical relevance, as compared to the classical two--sample log--rank test (see \cite{Mantel:1966, Peto:1972}).  Asymptotically (i.e. for sufficiently large sample size) the classical two--sample log-rank test is known to be the optimal test under proportional hazards (PH) alternatives. It is thus of particular interest to compare performance of both tests under PH alternatives.

In our simulations, patients were assumed to enter the trial uniformly between year $0$ and year $a$ with overall accrual rate of $r=100$ per year.
Accordingly, the calendar times of entry were generated according to a uniform distribution $Y_{x,i} \sim {\cal{U}}(0,a)$ on $[0,a]$. 
Patients were allocated equally to both treatment arms $A$ and $B$ (allocation ratio $\pi =1$), corresponding to an annual accrual rate of $50$ patients per group.
After the end of the accrual period, patients were assumed to be followed up for further $f=3$ years, while assuming no loss to follow--up. 
Accordingly, we set $C_{x,i}\coloneqq a+f-Y_{x,i}\sim {\cal{U}}(f,a+f)$. 
Survival times $T_{A,i}$ in the control intervention group $A$ were generated acc. to a Weibull distribution $\Lambda_A(s)\coloneqq - \log(S_1) \cdot t^{\kappa}$ with prefixed shape parameter $\kappa$ and 1-year survival rate $S_{T_A}(1)=S_1$. To implement the PH condition, survival times $T_{B,i}$ in the experimental intervention group $B$ were generated according to a Weibull distribution with $\Lambda_B(s)\coloneqq \omega \cdot \Lambda_A(s)$, where $\omega$ is the true hazard ratio. The true hazard ratio $\omega$ has to be distinguished from the expected hazard ratio $\omega_0$, which defines the planning alternative $K_0: \Lambda_B = \omega_0 \cdot \Lambda_A$ underlying sample size calculation.

The classical two--sample log--rank test serves as reference. Sample size $n$ of the virtual trials was thus calculated as follows:
In a first step, we used Schoenfeld's formula from \cite{Schoenfeld:1981} to calculate the required number of events $d$ for the two--sample log--rank test to achieve a power of $1- \beta$ under the planning alternative $K_0: \Lambda_B = \omega_0 \cdot \Lambda_A$ for allocated two--sided significance level $\alpha$. The expected number of events under the planning alternative $K_0: \Lambda_B = \omega_0 \cdot \Lambda_A$ by calendar time $a+f$ is $E(d_A)=r/2 \cdot \int_f^{a+f}[1-(S_1)^{t^{\kappa}}] dt$ in the standard treatment group A, and $E(d_B)=r/2 \cdot \int_f^{a+f}[1-(S_1)^{\omega_0 \cdot t^{\kappa}}] dt$ in the experimental treatment group B. Solving the condition $d=E(d_A)+E(d_B)$ for the  indeterminate $a$ yields the required length of the accrual period. The required total sample size is $n\coloneqq r \cdot a$ (i.e. $n/2$ per treatment group).

To cover scenarios of larger and smaller sample sizes, we let the expected hazard ratio $\omega_0$ range in the set  $\{0.5, 0.67, 0.8\}$.
To study the impact of different shapes of the survival distribution, we considered different values for the Weibull shape parameter from the interval $[0.1,5]$.
To study the impact of the event rate, we chose a reference arm 1-year survival rate $S_1$ of $0.5$ (Table \ref{tab:IntEventRate}), $0.8$ (see Appendix C) and $0.2$ (see Appendix D).

\begin{table}[!ht]
		\small
		\caption{\bf Comparison of empirical type I and II errors of new procedure and two-sample log-rank test}
		\begin{tabular}{cc|ccccc|ccccc}
			\hline
			& & \multicolumn{5}{c|}{Scenario 1} & \multicolumn{5}{c}{Scenario 2} \\[0.14cm]
			$\kappa$  & $\omega_0$ & $n$ & $\alpha_{new}$ & $\alpha_{LR}$ & $1-\beta_{new}$ & $1-\beta_{LR}$ & $n'$ & $\alpha_{new}$ & $\alpha_{LR}$ & $1-\beta_{new}$ & $1-\beta_{LR}$  \\
			\hline
			0.1  &   0.50    &   150   &   0.048   &   0.049   &   0.802   &   0.798  &  123    &   0.046     &   0.050     &   0.708   &   0.710  \\
			0.1  &   0.67    &   402   &   0.049   &   0.049   &   0.805   &   0.799  &  359    &   0.047     &   0.048     &   0.757   &   0.752  \\
			0.1  &   0.80    &  1180   &   0.045   &   0.045   &   0.803   &   0.798  & 1116    &   0.044     &   0.045     &   0.781   &   0.776  \\ [0.14cm]
			0.25 &   0.50    &   122   &   0.048   &   0.051   &   0.803   &   0.800  &  111    &   0.045     &   0.049     &   0.717   &   0.720  \\
			0.25 &   0.67    &   346   &   0.051   &   0.050   &   0.809   &   0.802  &  319    &   0.050     &   0.050     &   0.772   &   0.764  \\
			0.25 &   0.80    &   986   &   0.049   &   0.048   &   0.814   &   0.806  &  957    &   0.049     &   0.049     &   0.798   &   0.790  \\[0.14cm]
			0.5  &   0.50    &   110   &   0.047   &   0.051   &   0.809   &   0.800  &   96    &   0.045     &   0.050     &   0.749   &   0.743  \\
			0.5  &   0.67    &   284   &   0.049   &   0.050   &   0.815   &   0.802  &  270    &   0.051     &   0.051     &   0.795   &   0.780  \\
			0.5  &   0.80    &   798   &   0.050   &   0.050   &   0.818   &   0.803  &  789    &   0.051     &   0.051     &   0.811   &   0.797  \\[0.14cm]
			0.75 &   0.50    &   94    &   0.049   &   0.051   &   0.811   &   0.801  &   84    &   0.049     &   0.056     &   0.765   &   0.756  \\
			0.75 &   0.67    &   244   &   0.050   &   0.051   &   0.816   &   0.796  &  236    &   0.050     &   0.050     &   0.802   &   0.782  \\
			0.75 &   0.80    &   702   &   0.053   &   0.050   &   0.826   &   0.802  &  698    &   0.053     &   0.049     &   0.823   &   0.799  \\[0.14cm]
			1    &   0.50    &   82    &   0.050   &   0.056   &   0.810   &   0.799  &   76    &   0.048     &   0.056     &   0.778   &   0.766  \\
			1    &   0.67    &   220   &   0.051   &   0.052   &   0.821   &   0.798  &  216    &   0.051     &   0.052     &   0.815   &   0.792  \\
			1    &   0.80    &   658   &   0.051   &   0.050   &   0.826   &   0.803  &  657    &   0.050     &   0.051     &   0.823   &   0.801  \\[0.14cm]
			1.25 &   0.50    &   76    &   0.037   &   0.059   &   0.803   &   0.801  &   70    &   0.036     &   0.058     &   0.758   &   0.768  \\
			1.25 &   0.67    &   208   &   0.051   &   0.054   &   0.817   &   0.803  &  203    &   0.047     &   0.054     &   0.758   &   0.768  \\
			1.25 &   0.80    &   640   &   0.051   &   0.052   &   0.816   &   0.800  &  639    &   0.050     &   0.052     &   0.811   &   0.800  \\[0.14cm]
			1.5  &   0.50    &   72    &   0.029   &   0.059   &   0.731   &   0.801  &   67    &   0.024     &   0.059     &   0.616   &   0.764  \\
			1.5  &   0.67    &   200   &   0.045   &   0.055   &   0.799   &   0.799  &  198    &   0.045     &   0.055     &   0.796   &   0.792  \\
			1.5  &   0.80    &   634   &   0.051   &   0.053   &   0.814   &   0.800  &  633    &   0.050     &   0.052     &   0.809   &   0.792   \\[0.14cm]
			2    &   0.50    &   68    &   0.027   &   0.058   &   0.496   &   0.793  &   66	  &   0.026 	  &   0.059     &   0.464   &   0.779   \\
			2    &   0.67    &   198   &   0.042   &   0.056   &   0.757   &   0.797  &  196    &   0.041     &   0.055     &   0.748   &   0.792   \\
			2    &   0.80    &   632   &   0.053   &   0.052   &   0.811   &   0.799  &  631    &   0.050     &   0.052     &   0.806   &   0.798   \\[0.14cm]
			5    &   0.50    &   66    &   0.026   &   0.059   &   0.402   &   0.779  &   66    &   0.026     &   0.059     &   0.402   &   0.779   \\
			5    &   0.67    &   196   &   0.041   &   0.055   &   0.742   &   0.792  &  195    &   0.037     &   0.055     &   0.724   &   0.788    \\
			5    &   0.80    &   632   &   0.053   &   0.052   &   0.811   &   0.799  &  631    &   0.050     &   0.052     &   0.805   &   0.798    \\[0.14cm]
			\hline
		\end{tabular}
		\begin{flushleft} Empirical type I error rates ($\alpha_{new}$ and $\alpha_{LR}$) and powers ($1-\beta_{new}$ and $1-\beta_{LR}$) for the new test and for the classical two--sample log--rank test, respectively, under proportional hazards alternatives for Weibull distributed survival times with shape parameter $\kappa$ and 1--year survival rate $S_1=0.5$ in the control arm. Theoretical two--sided significance level: $5 \%$. Underlying total sample size $n$ (or $n'$) in Scenario 1 (or Scenario 2) calculated to achieve a theoretical power of $80 \%$  under the planning alternative $H_1:\Lambda_B = \omega_0 \cdot \Lambda_A$for the classical log--rank test using Schoenfeld's formula (or for the new test using formula (\ref{Sec_04_eq03})). 
		\end{flushleft}
		\label{tab:IntEventRate}
\end{table}

For each parameter constellation, we generated 10000 samples of size $n$ to which we applied both the new test as well as the classical two--sample log--rank test. We finally also used formula (\ref{Sec_04_eq02}) to calculate the sample size $n'$ such that our new test achieves a power of $1 - \beta$ under planning alternative $K_0: \Lambda_B = \omega_0 \cdot \Lambda_A$ for allocated two--sided significance level of $5\%$, and then repeated above simulations based on a total sample size of $n'$ instead of $n$. Reported in Tables 1--3 are the empirical type I and type II error rates for each parameter constellation and test based on a sample size of $n$ (Scenario 1) or $n'$ (Scenario 2).

%======================================================================================================================
\subsection{Results of the Main Setting (Table \ref{tab:IntEventRate}, Scenario 1)}\label{Sec_05_02}
%======================================================================================================================

Reassuringly, for large sample sizes ($\omega_0 = 0.8$), both tests preserve the desired significance level and achieve similar power levels close to the desired $80\%$ for all shape parameter values $\kappa$. On closer inspection, one notices that both tests tend to be conservative for small values of $\kappa$, and slightly anti--conservative for larger values of $\kappa$ on an acceptable degree (empirical type I error ranging between $4.5\%$ for $\kappa=0.1$ and $5.3\%$ for $\kappa=5$ and $\omega_0 = 0.8$). For the classical two--sample log--rank test, this effect is overlapped by a general tendency to anti--conservativeness when sample size is small ($\omega = 0.5$), resulting in an empirical type I error up to $5.9\%$ for the classical two--sample log--rank test when $\kappa=5$ and $\omega=0.5$.

For shape parameters $\kappa \leq 1$, both test perform similarly well with empirical type I error rate close to $5\%$. Interestingly, the new test even surpasses the classical two--sample log--rank test regarding power performance when $\kappa \leq 1$. This effect is most pronounced for exponentially distributed survival time  ($\kappa = 1$), when the new test achieves a power up to $83\%$ as compared to $80\%$ for the classical two--sample log--rank test. For shape parameter close to the exponential distribution $\kappa \approx 1$, the new test is observed to show even better type I error rate control than the classical two--sample log--rank test when sample size is small ($\omega = 0.5$).

For the extreme scenario of large shape parameters $\kappa \geq 2$ in combination with small sample size $\omega = 0.5$, however, the new test is observed to become quite conservative with profound loss in power ($40\%$ instead of $80\%$ for $\kappa = 5$ and $\omega = 0.5$). This is due to the fact that the new test requires estimation of the control arm cumulative hazard function, which seems to fail when sample size of the control arm is small and at the same time early events are rare ($n_A=34$ for $\kappa = 5$ and $\omega = 0.5$). In contrast, the classical two--sample log--rank test maintains power also in these extreme scenarios, with a tendency towards anti--conservativeness, though. 

This behavior of both tests is consistently observed amongst scenarios with different event rates (see the tables in Appendices C and D).

%======================================================================================================================
\section{Case Study}\label{sec:case_study}
%======================================================================================================================

As seen in the preceding simulations, the type I error of the classical one-sample log-rank test always exceeds the nominal type I error level if the sampling variability of the reference curve is not taken into account. However, the magnitude of this excess depends on the data from the reference cohort as well as the sample size in the new, experimental cohort.\\
The only difference  between the test statistic of the classical one-sample log-rank test

\begin{equation*}
Z_{\text{OSLR}} \coloneqq \frac{\widehat{M}_0(\infty)}{\widehat{\Sigma}_{\text{OSLR}}(\infty)} \qquad \text{with} \qquad {\widehat{\Sigma}^2_{\text{OSLR}}(s)}\coloneqq n_B^{-1} N_B(s)
\end{equation*}
and the new test is the denominator. Let $R\coloneqq {\widehat{\Sigma}_{\text{OSLR}}(\infty)}/\widehat{\Sigma}(\infty)$ denote the ratio of the standardisations without and with consideration of the sampling variability. The expected level of a two-sided classical one-sample log-rank test with nominal level $\alpha$ neglecting the sampling variability is then given by $E[2\cdot\Phi\left( \sqrt{R} \cdot z_{\frac{\alpha}{2}} \right)]$ which can be approximated by $2\cdot\Phi\left( E[\sqrt{R}] \cdot z_{\frac{\alpha}{2}} \right)$. Analogously, $E[\sqrt{R}]$ can be approximated via a first-order Taylor expansion by

\begin{equation*}
\sqrt{E[{\widehat{\Sigma}_{\text{OSLR}}(\infty)}]/E[\widehat{\Sigma}(\infty)]}
\end{equation*}
which is now a quantity we can estimate from given historical control data and design parameters of a trial.\\
From the computations in \cite{Wu:2015} we get

\begin{equation}\label{eq:expectation_classical_variance}
E[{\widehat{\Sigma}^2_{\text{OSLR}}(\infty)}]=\int_0^{\infty} F_{T_B}(u) dF_{C_B}(u).
\end{equation}
After another approximation and some computations (see Appendix B), we also get 

\begin{align*}
E[\widehat{\Sigma}^2(\infty)]\approx &\int_0^{\infty} F_{T_B}(u) dF_{C_B}(u) \\&+ 2\pi \cdot\left( \int_0^{\infty} \widehat{\sigma}_A(u)S_{T_B}^2(u)S_{C_B}(u) dF{C_B}(u) + \int_0^{\infty} \widehat{\sigma}_A(u)S_{T_B}(u)S_{C_B}^2(u) dF_{T_B}(u)\right)
\end{align*}
Under the null hypothesis, this can be estimated by plugging in Kaplan-Meier estimates from the control group for $F_{T_B}$ resp. $S_{T_B}$. For a given historical control group, these formulas can now be used to compute the type I error inflation when sampling variability is not taken into account. Of course, the treatment group allocation ratio $\pi$ is essential for the extent of this inflation.\\
We will now illustrate the influence of basic design parameters on the type I error inflation with a practical example. We employ the setting of the Mayo Clinical trial in primary biliary cirrhosis of the liver (PBC), which is a rare but fatal chronic disease whose cause is still unknown (see \cite{Fleming:2005}). In this double-blinded randomized trial the drug D-penicillamine (DPCA) was compared with a placebo. The study data is publicly available via the survival package in R \cite{Therneau:2020, R}.\\

\begin{figure}[!h]
	\centering
	\includegraphics[width=\textwidth]{"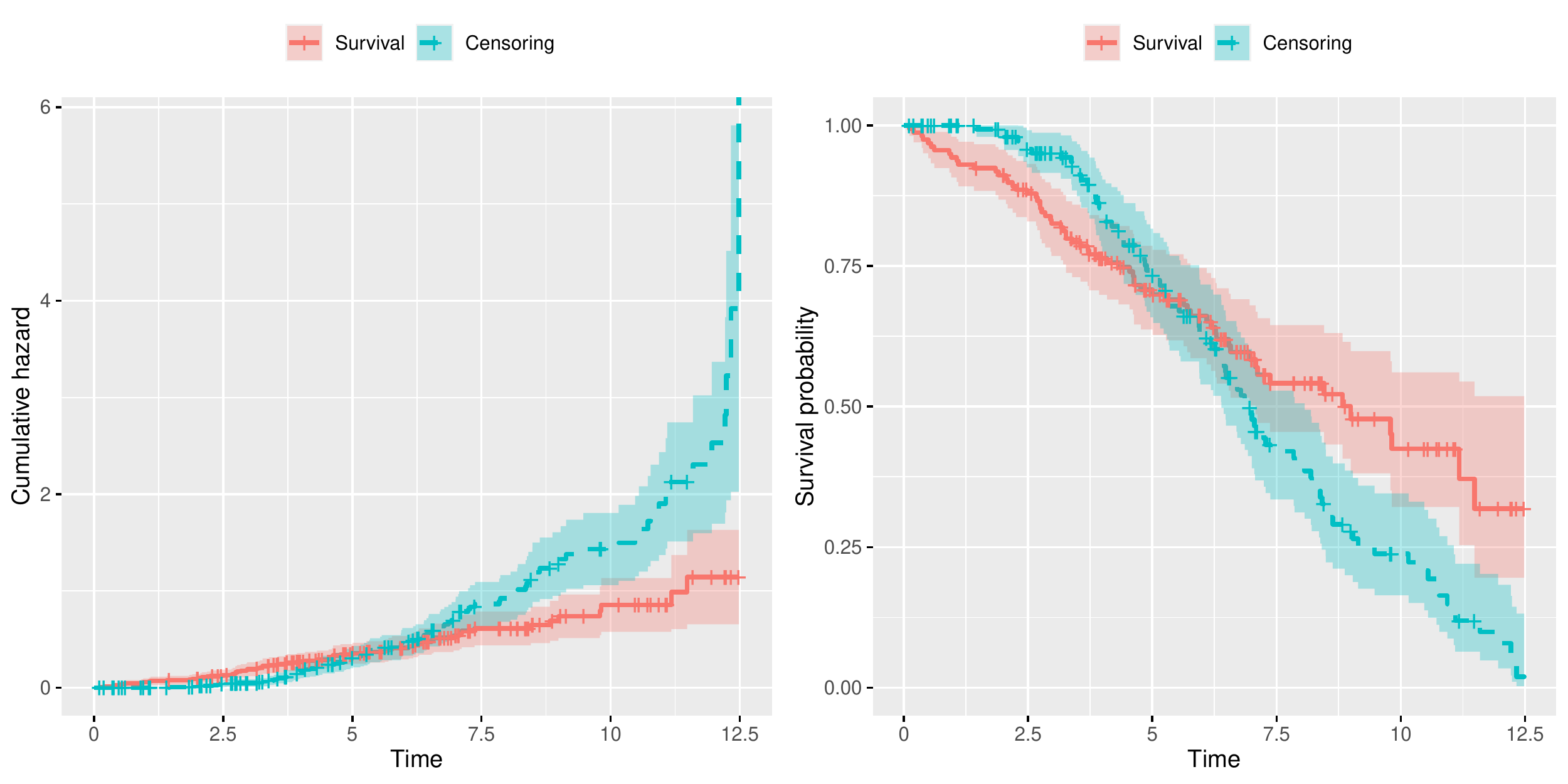"}
	\caption{{\bf Distribution of survival and censoring variable.}
		Distribution of overall survival and censoring in the cohort treated with DPCA of the Mayo Clinical trial in primary biliary cirrhosis. Left: Cumulative hazards according to the Nelson-Aalen estimator. Right: Survival distributions accoring to the Kaplan-Meier estimator}
	\label{fig:control_cohort}
\end{figure}

\begin{figure}[!h]
	\centering
	\includegraphics[width=\textwidth]{"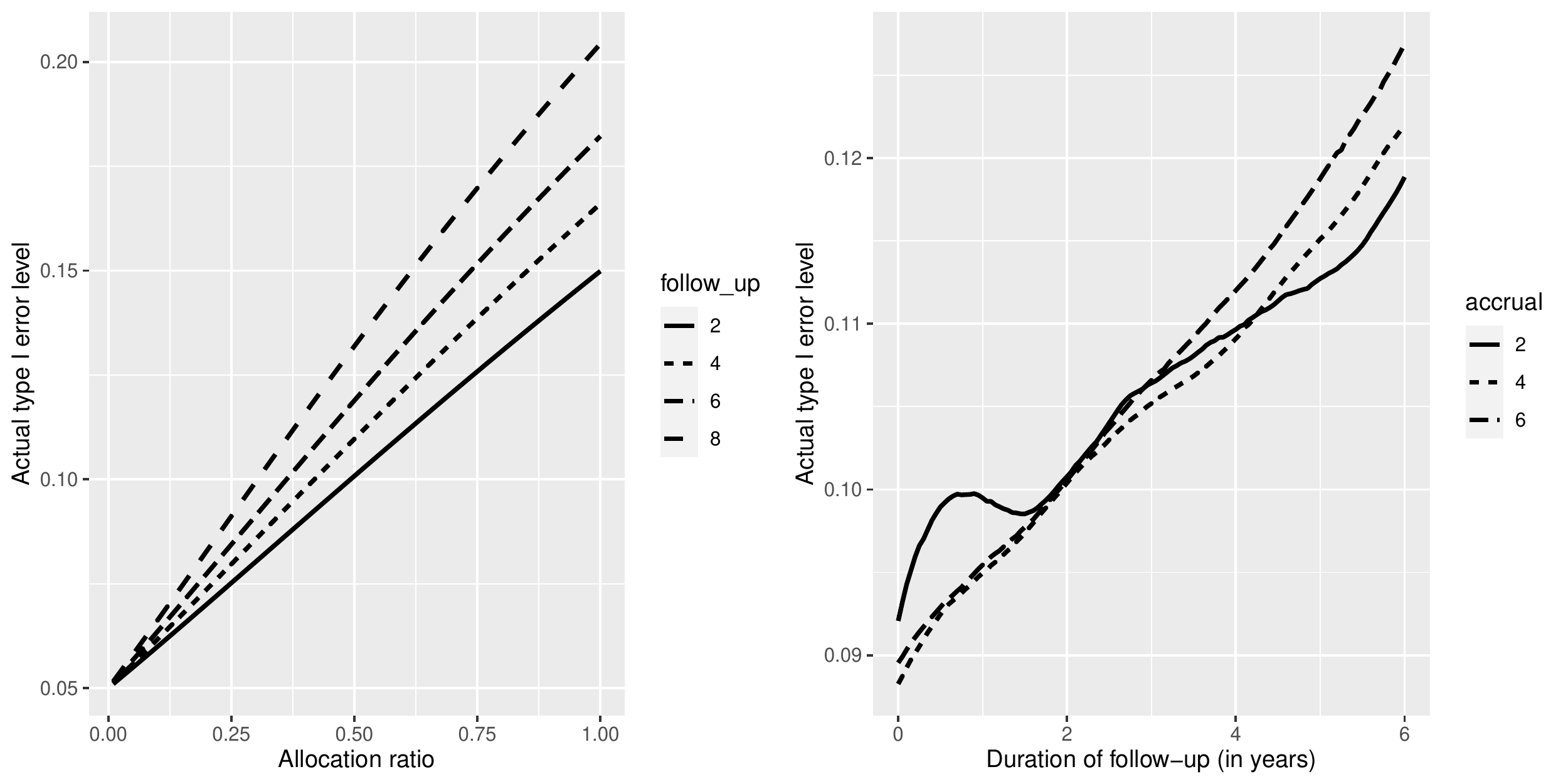"}
	\caption{{\bf Type I error inflation.}
		Actual type I error levels of the classical one-sample log-rank test. Left: Variation of the allocation ratio with fixed accrual duration and four different durations of the follow-up period. Right: Variation of the length of the follow-up period for a fixed allocation ratio and three different durations of the accrual period.}
	\label{fig:type_1_errors}
\end{figure}

Among the 158 patients of the cohort treated with DPCA, 65 died during the trial. The Kaplan-Meier survival curve of these patients can be found in Fig \ref{fig:control_cohort}. The time scale is given in years. There, we also display the empirical distribution of the censoring variable $C$ in this cohort. As we will see below, this distribution also plays a substantial role for our computations here. We now suppose, that a new treatment becomes available and the data from this trial shall be used to compare the survival under this treatment to the survival under treatment with DPCA. This shall be accomplished in a trial in which patients are recruited uniformly over a accrual period of length $a$ and followed-up in an additional period of length $f$. The allocation ratio will again be denoted by $\pi$. If one cannot find a suitable parametric model to be fitted to the data, the Kaplan-Meier resp. Nelson-Aalen estimates (see Fig \ref{fig:control_cohort}) are employed as reference curves for the one-sample log-rank test.\\
Similar to our first simulation study (see \nameref{Sec_06}), we  investigate the influence of the allocation ratio on the inflation of the type I error level in the first part of our study. We choose $\pi\in \{0.01, 0.02, 0.03,\dots, 1\}$, $a=2$ and $f\in\{2, 4, 6, 8\}$. The results in terms of the actual type I error level of the one-sample log-rank test can be found on the left hand side of Fig \ref{fig:type_1_errors}. For any fixed $f$, the actual type I error level increases nearly linearly in the range of allocation ratios considered here. So, as a rule of thumb, each additional trial patient raises the level by a fixed number of percentage points. This number however seems to depend on the length of the follow-up, where a longer duration of the follow-up period leads to steeper increases.\\
In the second part of this case study, we take a closer look at the role of the trial duration. As already seen in the first part, longer trials lead to a larger inflation of the error levels. To analyse this dependence, we now choose $\pi=0.5$, $a\in\{2,4,6\}$ and $f\in\{0, 0.05, 0.1,\dots, 6\}$. The results can be found on the right hand side of Fig \ref{fig:type_1_errors}. As we can see, trials with a longer total duration ($a+f$) tend to lead to a higher type I error inflation. This effect is most substantial if the total trial duration is close to the longest observation in the reference data set which amounts to about 12.5 years. In this case, the testing procedure needs to utilize parts of the Nelson-Aalen estimator which are affected by a high amount of variability because of the high number of censored observations. However, the inflated type I error neither behaves completely monotonically w.r.t. the accrual duration $a$ nor the follow-up duration $f$. Even if the variance of the classical one-sample log-rank test and the additional variance which is due to the sampling variability (see appendix A) increase monotonically in $a$ and $f$, the ratio $R$ can increase if the increase of the former is steeper than the increase of the latter. Nevertheless, there is a clear tendency towards a larger inflation of the type I error if either $a$ or $f$ increases.

%======================================================================================================================
\section{Discussion}\label{sec:discussion}
%======================================================================================================================

Traditional one--sample log--rank tests compare the survival function of an experimental treatment to a prefixed reference survival curve, which typically represents the expected survival under standard of care. Choice of the reference survival curve is typically based on historic data on standard therapy and thus prone to statistical error. Nevertheless, traditional one--sample log--rank tests do not account for this variance of the reference curve estimator. Here we study and propose a non--parametric one--sample log--rank test that explicitly accounts for sampling variability of the reference curve. 

The new test may also be interpreted as two--sample test for survival distributions, while inheriting the interpretability from the underlying one--sample log--rank test. Admittedly, our simulations suggest that it may be advisable to compare the data of a historical control cohort with the new data in a single-arm Phase II trial via the two-sample log-rank test if one wants to account for sampling variability of the reference curve or in case of allocation ratios close to 1. Nevertheless, in Phase II settings with fast events (Weibull shape parameter $\kappa \leq 1$), our simulations reveal the potential of the new test to outperform the classical two--sample log--rank test even under PH alternatives. A non-consideration of the sampling variability leads to an inflation of the type I error rate. The extent of this inflation depends in particular on the size of the control cohort. A major objective of this work was to investigate how large this control must be chosen so that the type I error inflation remains within an acceptable range. In this regard, our simulations support that the classical one-sample log-rank test is adequate if the historical control cohort is at least about 10 times larger than the new cohort ($\pi \leq 1/10$) and the maximum follow-up in the new trial is reasonably small in view of the follow-up duration in the historic cohort (see \nameref{sec:case_study}).

Conceptually, the proposed new test also sheds light on a general strategy for lifting existing methodology for single--arm survival trials to a randomized, multi--arm setting. This might be of interest for designing confirmatory survival trials with interim analyses. Performance of interim analyses in clinical trials is of ethical and economic interest. On the one hand, interim analyses enable faster decisions regarding rejection or acceptance of the underlying null hypothesis when the treatment effect is larger or smaller than initially expected. Moreover, interim analyses offer the possibility for data dependent modifications of the trial (e.g. sample size recalculation) in the case of new insights, thus increasing the prospects of the trial. Trial designs with interim analyses offering such kind of flexibility at full type I error rate control are commonly referred to as \emph{confirmatory adaptive designs} \cite{Bauer:1989, Bauer:1994}. Whereas methodology for confirmatory adaptive designs is well understood for trials with short--term endpoints as in \cite{BPB02, Hommel:2001}, subtle problems arise for adaptive survival trials. With standard methodology for group--sequential adaptive survival trials from \cite{Wassmer:2006}, the degree of flexibility is highly limited. For example, in a survival trial with primary endpoint \emph{overall survival (OS)}, essentially only interim information on the survival status of the patients may be used for design modifications (e.g. sample size recalculation). Further interim information, e.g. on progression status of the patients, must not be used for design modifications in these classical adaptive Phase III survival trials, because otherwise type I error rate inflation may occur (see \cite{Bauer:2004}). This situation is clinically unsatisfactory. If larger degree of flexibility is desired, the patient--wise separation approach as initially proposed by \cite{Jenkins:2011} has to be chosen which, however, either implies neglection of some part of the observed survival data in the test statistic or requires some worst--case adjustments resulting in a conservative design as shown in \cite{Magirr:2016}. Until today, no satisfactory methodology for adaptive Phase III survival trial exists, that offers larger flexibility while avoiding those problems involved with the patients--wise separation approach.
Recently, however, such methodology was proposed for single arm Phase II survival trials. In \cite{Danzer:2021b}, an adaptive one--sample log--rank test was suggested that allows the simultaneous use of several time--to--event endpoints for data--dependent design modifications, while  avoiding those problems involved with the patient--wise separation approach. In a similar way the common one--sample log--rank test was lifted to a two--sample setting in this paper, we expect that the multivariate adaptive one--sample log--rank test proposed by \cite{Danzer:2021b} may be lifted to a two--sample setting, thus solving an outstanding problem in the theory of adaptive design methodology. Implementation of this idea, however, is beyond the scope of this paper and will be contents of an upcoming paper. The objective of this paper is to provide methodology for accounting for sampling variability of the reference curve in classical one--sample log--rank tests, and to show  feasibility of the underlying lifting procedure regarding type I and type II error rate control.

\section{Acknowledgments}
The work of Moritz Fabian Danzer was funded by the German Science Foundation (Deutsche Forschungsgemeinschaft, DFG, grant number 413730122).

\newpage

%======================================================================================================================
\section*{Appendix A: Proof of Distributional Properties}\label{appA}
%======================================================================================================================

\textbf{Theorem 1.}
\textit{Let $s_0 > 0$ be given s.t. $S_{X_A}(s_0)=S_{T_A}(s_0)S_{C_A}(s_0)\eqqcolon p_0 > 0$ and assume 
	%the following conditions:
	%\begin{itemize}
	%\item[(i)] No two event indicators $N_{x,i}$ and $N_{x',j}$ with $x \neq x'$ or $i \neq j$ jump simultaneously.
	%\item[(ii)] There is an $n_0$ and a $c_0 \in (0,1)$ s.t. for all $n \geq n_0$ we have $Y_{A}(s_0) \geq c_0 n_A$ (i.e. for sufficiently large sample size there is a certain fraction of control arm patients surviving the whole trial period).
	%\item[(iii)]  Pointwise for each $0 \leq s \leq s_0$, we have $Y_x(s)/n_x \stackrel{{\cal{P}}}{\rightarrow} y_x(s)$ in probability as $n \to \infty$ for some strictly decreasing, deterministic function $y_x:[0,s_0] \longrightarrow [0,1]$.
	%\end{itemize}
	%Moreover assume 
	that the null hypothesis $H_0: \Lambda_A(s)=\Lambda_B(s)$ for all $0 \leq s \leq s_0$ is true. Set
	\begin{align}\label{appA_eq01}
	\begin{split}
	\widehat{M}_0(s) &\coloneqq n_B^{-1/2}\left[N_B(s) - \sum_{i \in {\cal{N}}_{B}} \widehat{\Lambda}_A(s \wedge X_{B,i})\right] \\
	\widehat{\Sigma}^2(s) &\coloneqq n_B^{-1} N_B(s) + n_B^{-1} n_A^{-1} \sum_{i,j \in {\cal{N}}_{B}} \widehat{\sigma}_A(s \wedge X_{B,i} \wedge X_{B,j})
	\end{split}
	\tag{A.1}
	\end{align}
	Then the following is true:
	\begin{itemize}
		\item $\widehat{M}_0|_{[0,s_0]}$ has asymptotically independent increments, i.e. for all $0 \leq s_1 \leq s_2 \leq s_0$ and sufficiently large sample size $n$, the random variables $\widehat{M}_0(s_1)$ and $\widehat{M}_0(s_2)-\widehat{M}_0(s_1)$ are approximately independent.
		\item Pointwise, for each $0 \leq s \leq s_0$, we have $\widehat{M}_0(s) \stackrel{\cal{D}}{\rightarrow} {\cal{N}}(0,\Sigma^2(s))$ as $n \to \infty$, where $\Sigma(s)\coloneqq  \emph{plim}_{n \to \infty} \widehat{\Sigma}(s) = \lim_{n \to \infty} E[\widehat{\Sigma}(s)]$ is the large sample limit of $\widehat{\Sigma}(s)$ (existing acc. to Lemma 1 below).
	\end{itemize}
}

{\small{\textsc{Proof.}}}
It is well known that $M_{x,i}(s)\coloneqq N_{x,i} - \int_0^s I(X_{x,i} \geq u) \lambda_x(u) du$ is a mean--zero ${\cal{F}}_s$--martingale with optional covariation $[M_{x,i}](s)\coloneqq N_{x,i}$. By independence of the summands, it follows that $M_{x}(s)\coloneqq N_{x} - \int_0^s Y_x(u) \lambda_x(u) du$ is a mean--zero ${\cal{F}}_s$--martingale with optional covariation $[M_{x}](s)\coloneqq N_{x}(s)$. In particular, for any left--continuous ${\cal{F}}_s$--adapted process $H(s)$, $M_H(s)\coloneqq\int_0^s H(u) dM_A(u)$ is a mean zero ${\cal{F}}_s$--martingale with optional covariation $\int_0^sH^2(u)dN_A(u)$. Choosing $H(u)\coloneqq J_A(u)/Y_A(u)$ we recover the well--known result that $M_{J_A/Y_A}(s)\coloneqq\int_0^s \frac{J_A(u)}{Y_A(u)} dM_A(u)$ is a mean zero ${\cal{F}}_s$--martingale with optional covariation $[M_{J_A/Y_A}](s)=\int_0^s \frac{J_A(u)}{Y_A^2(u)} dN_A(u)$. Notice that $M_{J_A/Y_A}(s)=\widehat{\Lambda}_A(s)-\Lambda_A^*(s)$, where $\widehat{\Lambda}_A(s)\coloneqq\int_0^s \frac{J_A(u)}{Y_A(u)} dN_A(u)$ is the Nelsen--Aalen estimate of $\Lambda_A(s)$, and $\Lambda_A^*(s)\coloneqq\int_0^s J_A(u)\lambda_A(u) du$. By independence of the treatment groups it follows that
\begin{align}\label{appA_eq02}
M_i(s)\coloneqq M_{B,i}(s) - M_{J_A/Y_A}(s)
\tag{A.2}
\end{align}
is a mean--zero ${\cal{F}}_s$--martingale with optional covariation $[M_{i}](s)\coloneqq [M_{B,i}](s) + [M_{J_A/Y_A}](s)$. Since $\tau_i\coloneqq X_{B,i}$ is an ${\cal{F}}_s$--stopping--time (see Lemma 2 below), we conclude from appeal to the \emph{optional stopping theorem} and \emph{compatibility of stopping with covariation} that the stopped process $M_i^{\tau_i}(s)\coloneqq M_i(s \wedge \tau_i)$ 
%\begin{align}\label{appA_eq03}
%M_i^{\tau_i}(s) = M_{B,i}^{\tau_i}(s) - M_{J_A/Y_A}^{\tau_i}(s) = M_{B,i}(s \wedge \tau_i) - M_{J_A/Y_A}(s \wedge \tau_i)
%\end{align}
is a mean--zero ${\cal{F}}_s$--martingale with
\begin{align}\label{appA_eq03}
\begin{split}
%&\bullet \quad M_i^{\tau_i}(s) = M_i^{\tau_i}(s \wedge \tau_i) = M_{B,i}(s \wedge \tau_i) - M_{J_A/Y_A}(s \wedge \tau_i), \\
&\bullet \quad [M_i^{\tau_i}](s)=[M_{i}](s \wedge \tau_i) = [M_{B,i}](s\wedge \tau_i) - [M_{J_A/Y_A}](s\wedge \tau_i), \\
&\bullet \quad [M_i^{\tau_i},M_j^{\tau_j}](s) = [M_{J_A/Y_A}](s\wedge \tau_i \wedge \tau_j) \quad \text{for } i \neq j.
\end{split}
\tag{A.3}
\end{align}
To see the last assertion in (\ref{appA_eq03}), use bilinearity of the covariation operator $[\cdot,\cdot]$ together with $[M_{B,i}^{\tau_i}M_{B,j}^{\tau_j}]=0$ and $[M_{B,i}^{\tau_i},M_{J_A/Y_A}^{\tau_j}]=0$ for $i \neq j$ by independence of the patients and treatment groups, where for any ${\cal{F}}_s$--adapted process $Q$ and any ${\cal{F}}_s$--stopping--time $\tau$ we use the common notation $Q^{\tau}(s)\coloneqq Q(s \wedge \tau)$. 

We are finally interested in the large sample properties of the mean zero ${\cal{F}}_s$--martingale
\begin{align}\label{appA_eq03b}
\begin{split}
\widehat{M}(s)\coloneqq n_B^{-1/2} \sum_{i \in {\cal{N}}_B} M_i^{\tau_i}(s).
\end{split}
\tag{A.4}
\end{align}
First notice that the jumpsize of $\widehat{M}$ is bounded by $2n_B^{-1/2}$ and thus vanishes in the large sample limit $n \to \infty$, because the jump sizes of $M_{B,i}$ and $M_{J_A/Y_A}$ are bounded by $1$, as no two event indicators $N_{x,i}$ and $N_{x',j}$ jump simultaneously a.s..
Making use of $N_{B,i}(s \wedge \tau_i) = N_{B,i}(s)$ (see Lemma 3 below) and noticing that $n_A \cdot [M_{J_A/Y_A}](s) \equiv \widehat{\sigma}_A(s)$ from (\ref{Sec_02_01_eq05}), some algebra shows that $\widehat{M}$ has optional covariation $[\widehat{M}](s)=\widehat{\Sigma}^2(s)$. So, by Lemma 1 below, $[\widehat{M}](s)$ converges pointwise in $s$ in probability to the strictly increasing, deterministic function $\Sigma^2(s) = \lim_{n \to \infty} E[\widehat{\Sigma}^2(s)]$.
All in all, we conclude from appeal to Rebolledo's martingale central limit theorem that $\widehat{M}$ converges  on $[0,s_0]$ in distribution to a mean zero Gaussian martingale $M^{(\infty)}$ with independent increments and variance function $\Sigma^2(s)$.

To finish the proof, it suffices to notice that the processes $\widehat{M}$ from (\ref{appA_eq03b}) and $\widehat{M}_0$ from (\ref{appA_eq01}) coincide in the limit when null hypothesis $H_0: \Lambda_A = \Lambda_B$ hold true.
\quad $\Box$ 
\ \\ \\
\textbf{Theorem 2.}
\textit{Fix $s_0>0$. Under the contiguous alternatives $\Lambda_B(\cdot)= \omega_n \Lambda_A(\cdot)$ with $\omega_n=\exp(-n^{-1/2}\gamma)$ for some $\gamma \geq 0$, the process $\widehat{M}_0$ defined in (\ref{appA_eq01}) converges on $[0,s_0]$ in distribution to a Gaussian process with independent increments, drift function $\mu(s)\coloneqq - \gamma \sqrt{\frac{\pi}{1+\pi}} \int_0^{\infty} F_{T_A}(s \wedge u) f_{C_A}(u) du$ and variance function $\Sigma^2(s)$ from (\ref{appA_eq11}).
}

{\small{\textsc{Proof.}}}
Under the contiguous alternatives, the difference between the mean--zero martingale $\widehat{M}$ from (\ref{appA_eq03b}) and $\widehat{M}_0$ is
\begin{align}\label{appA_eq14}
\begin{split}
\widehat{M}_0 - \widehat{M} 
= n_B^{-1/2} \sum_{i \in {\cal{N}}_B} [ \Lambda_B(s \wedge X_{B,i}) - \Lambda_A(s \wedge X_{B,i})]
= n_B^{1/2}[ 1 - \exp(n^{-1/2} \gamma) ] \cdot n_B^{-1} \sum_{i \in {\cal{N}}_B}  \Lambda_B(s \wedge X_{B,i}).
\end{split}
\tag{A.5}
\end{align}
As $n \to \infty$, the first factor converges to $- \gamma \sqrt{\frac{\pi}{1+\pi}}$. Since $\Lambda_B \approx \Lambda_A$ under the contiguous alternatives when $n$ is large, the second converges in probability to $E[\Lambda_A(s \wedge X_{A,1})]$ by law of large numbers as $n \to \infty$, which in turn coincides with $E[N_{A,1}(s)] = \int_0^{\infty} F_{T_A}(s \wedge u) f_{C_A}(u) du$  due to the martingale property of $M_{A,1}$. So the assertion follows from Theorem 1 and appeal to Slutsky's theorem.
\quad $\Box$
\ \\ \\
\textbf{Lemma 1.}
\textit{Let $s_0 > 0$ be given s.t. $S_{X_A}(s_0)=S_{T_A}(s_0)S_{C_A}(s_0)\eqqcolon p_0 > 0$. Let $\widehat{\Sigma}$ be the process defined in (\ref{appA_eq01}). Then there is a strictly increasing, deterministic function $\Sigma(s)$ with $\Sigma(0)=0$ such that pointwise for each $0 \leq s \leq s_0$, we have $\widehat{\Sigma}(s) \stackrel{{\cal{P}}}{\rightarrow} {\Sigma}(s)$ as $n \to \infty$. More specifically, $\Sigma(s) = \lim_{n \to \infty} E[\widehat{\Sigma}(s)]$.
}
\ \\ \\
{\small{\textsc{Proof.}}}
For this proof, we introduce the following abbreviations: 
\begin{align}
\begin{split}
\Theta&\coloneqq n_B^{-1} n_A^{-1} \sum_{i,j \in {\cal{N}}_{B}} \widehat{\sigma}_A(s \wedge X_{B,i} \wedge X_{B,j}), \\ 
\Psi_{N_A}(s)&\coloneqq \int_0^s \frac{J_A(u)}{Y_A^2(u)} dN_A(u), \\
\Psi_{M_A}(s)&\coloneqq \int_0^s \frac{J_A(u)}{Y_A^2(u)} dM_A(u), \\
\Psi_{\Lambda_A}(s)&\coloneqq \int_0^s \frac{J_A(u)}{Y_A(u)} \lambda_A(u) du.
\end{split}
\tag{A.6}
\end{align}
Since $n_B^{-1} N_B(s) \stackrel{{\cal{P}}}{\rightarrow} E[N_{B,1}(s)]$ as $n \to \infty$ by law of large numbers, it remains to prove that $\Theta \stackrel{{\cal{P}}}{\rightarrow} \lim_{n \to \infty} E[\Theta]$ as $n \to \infty$. The proof decomposes into several steps.
From appeal to triangle inequality we conclude that for any $\varepsilon > 0$
\begin{align}\label{appA_eq04}
\begin{split}
P\left(|\Theta - \lim_{n \to \infty} E[\Theta]| \geq 2 \varepsilon \right) 
\leq  P\left(|\lim_{n \to \infty} E[\Theta] - E[\Theta]| \geq \varepsilon \right) + P\left(|\Theta - E[\Theta]| \geq \varepsilon \right) .
\end{split}
\tag{A.7}
\end{align}
By Lemma 6, $\lim_{n \to \infty} E[\Theta]$ exists, i.e. the first summand on the right hand side of equation (\ref{appA_eq04}) vanishes in the limit $n \to \infty$. By conditioning on the outcomes in group B, the second summand can be rewritten as
\begin{align*}
P\left(|\Theta - E[\Theta]| \geq \varepsilon \right) &= E[P\left(|\Theta - E[\Theta]| \geq \varepsilon \right | X_B)]\\
&=\int_{[0,\infty)^{n_B}} P\left(|\Theta - E[\Theta]| \geq \varepsilon \right | X_B=x_B) \cdot f_{X_B}(x_B) dx_B 
\end{align*}
where $X_B=(X_{B,1}, \dots, X_{B,n_B})$ and $f_{X_B}=\prod_{i=1}^{n_B} f_{X_{B,1}}(x_{B,i})$ as the observations from group B are independent and identically distributed. Now, if $P\left(|\Theta - E[\Theta]| \geq \varepsilon \right | X_B=x_B)\leq c(n_A)$ for some function $c$ which does only depend on $n_A$ and not on $x_B$ with $c(n_A) \to 0$ as $n_A \to \infty$, we also have $P\left(|\Theta - E[\Theta]| \geq \varepsilon \right)$ as $n \to \infty$.\\
By Chebyshev's inequality, the Cauchy-Schwarz inequality and since $\Psi_{M_A}(s) = \Psi_{N_A}(s) - \Psi_{\Lambda_A}(s)$ we have for any $x_B\in [0, \infty)^{n_B}$ and any $\varepsilon > 0$
\begin{align}\label{appA_eq05}
\begin{split}
P(|\Theta - E[\Theta]| \geq \varepsilon|X_B = x_B) 
&\leq \frac{\text{Var}[\Theta|X_B = x_B]}{\varepsilon^2}\\
&= \text{Var}[n_B^{-1} n_A^{-1} \sum_{i,j \in \mathcal{N_B}} \widehat{\sigma}_A(s \wedge x_{B,i} \wedge x_{B,j})]\\
&\leq \frac{n_B^2}{n_A^2} \cdot \sum_{i,j \in \mathcal{N}_B} \text{Var} \left[\int_0^{s \wedge x_{B,i} \wedge x_{B,j}} n_A \cdot \frac{J_A(u)}{Y_A^2(u)} dN_A(u) \right]\\
&\leq \pi^2 \sup_{s^{\star} \in [0,s_0]} \text{Var} \left[\int_0^{s^{\star}} n_A \cdot \frac{J_A(u)}{Y_A^2(u)} dN_A(u) \right]\\
&\leq \pi^2 \left( \sup_{s^{\star} \in [0,s_0]} \sqrt{\text{Var}[n_A \cdot \Psi_{M_A}(s^{\star})]} + \sup_{s^{\star} \in [0,s_0]} \sqrt{\text{Var}[n_A \cdot \Psi_{\Lambda_A}(s^{\star})]} \right)
\end{split}
\tag{A.8}
\end{align}
where $\pi=n_B/n_A$ is the prefixed treatment group allocation ratio. Notice that the third inequality holds because $x_{B}$ is a fixed value and not a random variable. To finish the proof, we show that $\sup_{s^{\star} \in [0,s_0]} \text{Var}[n_A \cdot \Psi_{M_A}(s^{\star})] \to 0$ (Step I) and $\sup_{s^{\star} \in [0,s_0]} \text{Var}[n_A \cdot \Psi_{\Lambda_A}(s^{\star})] \to 0$ (Step II) as $n_A \to \infty$.

\emph{Proof of Step I}: 
\begin{align*}
\text{Var}\left[n_a \cdot \Psi_{\Lambda_A}(s)\right] & = n_A^2 \cdot \text{Var}\left[ \int_0^s \frac{J_A(u)}{Y_A(u)} \right]\\
=&n_A^2 \cdot \int_0^s \int_0^s \lambda_A(u)\lambda_A(v)\cdot \underbrace{\text{Cov}\left[\frac{J_A(u)}{Y_A(u)}, \frac{J_A(v)}{Y_A(v)}\right]}_{\leq \sup_{s^{\star} \in [0, s_0]} \text{Var}\left[ \frac{J_A(s^{\star})}{Y_A(s^{\star})} \right]} du dv
\end{align*}
For any $s \in[0,s_0]$, we also have
\begin{align*}
\text{Var} \left[ \frac{J_A(s)}{Y_A(s)} \right]&=E\left[\left( \frac{J_A(s)}{Y_A(s)} - E\left[ \frac{J_A(s)}{Y_A(s)}\right] \right)^2\right]\\
&\leq E\left[\left( \frac{J_A(s)}{Y_A(s)} - \frac{1}{n_A\cdot S_{X_A}(s)} \right)^2\right]\\
&\leq \frac{1}{n_A^2} \left|E\left[ \frac{n_A^2\cdot J_A(s)}{Y^2_A(s)} - \frac{1}{S_{X_A}(s)^2} \right]\right| + \frac{2}{n_A^2 \cdot S_{X_A}(s)} \left|E\left[ \frac{n_A\cdot J_A(s)}{Y_A(s)} - \frac{1}{S_{X_A}(s)} \right]\right|
\end{align*}
For both summands we can apply the results from Lemma 4.2 (i) of \cite{McKeague:1990} as the probability that $Y_A(s)=0$ goes to zero uniformly on $[0,s_0]$ to get
\begin{equation*}
\text{Var} \left[ \frac{J_A(s)}{Y_A(s)} \right] \leq \frac{K}{n_A^3} + \frac{2K}{n_A^3\cdot p_0^3}
\end{equation*}
We can finally plug this estimate in to obtain
\begin{equation*}
\sup_{s^{\star} \in [0,s_0]} \text{Var}\left[n_A \cdot \Psi_{\Lambda_A}(s^{\star}) \right] \leq \Lambda_A^2(s_0) \cdot \left( \frac{K}{n_A} + \frac{2K}{n_A\cdot p_0^3} \right)
\end{equation*}
which goes to zero on $[0,s_0]$ as $n_A \to \infty$.\\
\emph{Proof of Step II}: 
$\Psi_{M_A}(s)$ is a mean zero ${\cal{F}}_s$--martingale since $M_A(s)$ is so. Consequently, $\Psi_{M_A}^2(s) - [\Psi_{M_A}](s)$ is a mean zero ${\cal{F}}_s$--martingale, where $[\Psi_{M_A}](s) = \int_0^s \frac{J_A(u)}{Y_A^4(u)} dN_A(u)$ is the optional covariation of $\Psi_{M_A}(s)$. Thus
\begin{align}\label{appA_eq07}
\begin{split}
\text{Var}\left[ \Psi_{M_A}(s) \right]
&= E \left[ \Psi_{M_A}^2(s) \right]
= E \left[ [\Psi_{M_A}](s) \right]
= E \left[ \int_0^s \frac{J_A(u)}{Y_A^4(u)} dN_A(u) \right] \\
&= E \left[ \int_0^s \frac{J_A(u)}{Y_A^3(u)} \lambda_A(u) du \right]
=\int_0^s E\left[\frac{J_A(u)}{Y^3_A(u)} \lambda_A(u) du \right] \\
&\leq \int_0^s \frac{4^3}{n_A^3 \cdot S^3_{X_A}(u)} \lambda_A(u) du
\leq \frac{4^3}{n_A^3 \cdot p_0^3} \cdot \Lambda_A(s_0).
\end{split}
\tag{A.9}
\end{align}
where the first inequality from the third row follows from Lemma 1 from \cite{Aalen:1976}. Those inequalities hold for all $s\leq s_0$. We thus conclude that
\begin{align}\label{appA_eq08}
\begin{split}
\sup_{s^{\star} \in [0,s_0]}\text{Var}\left[ n_A \cdot \Psi_{M_A}(s^{\star}) \right]
\leq n_A^{-1}   \Lambda_A(s_0) \cdot \frac{4^3}{p_0^3}.
\end{split}
\tag{A.10}
\end{align}
for any $s\leq s_0$ which finishes the proof
\quad $\Box$
\ \\ \\
\textbf{Lemma 2.}
\textit{$\tau_i\coloneqq X_{B,i}$ is an ${\cal{F}}_s$--stopping--time. 
}

{\small{\textsc{Proof.}}}
$\{\tau_i > s\} = \{ T_{B,i} \wedge C_{B,i} > s\} = \{ T_{B,i} > s\} \cap \{ C_{B,i} > s\} \in {\cal{F}}_s$.
\quad $\Box$
\ \\ \\
\textbf{Lemma 3.}
\textit{For any two random variables $S$ and $T$ we have $I(T \leq S) = I(T \leq S \wedge T)$. In particular, $N_{B,i}(s \wedge \tau_i) = N_{B,i}(s)$ for the ${\cal{F}}_s$--stopping--time $\tau_i\coloneqq X_{B,i}$.
}

{\small{\textsc{Proof.}}}
$\{ T \leq T \wedge S \} = \{  T \leq T \wedge S, T \leq S \} \cup \{  T \leq T \wedge S, T > S \} = \{  T \leq T, T \leq S \} \cup \{  T \leq S, T > S \} = \{ T \leq S \}$.
\quad $\Box$
\ \\ \\
\textbf{Lemma 4.}
\textit{Consider $\widehat{\sigma}_x(s)$ from Eq. (\ref{Sec_02_01_eq05}). Then in probability as $n \to \infty$
	\begin{align}\label{appA_eq09}
	\widehat{\sigma}_x(s) \stackrel{{\cal{P}}}{\rightarrow} {\sigma}_x(s) \coloneqq \int_0^s \frac{\lambda_x(u)}{S_{T_x}(u) \cdot S_{C_x}(u)}  du,
	\tag{A.11}
	\end{align}
	where $S_{T_x}$ ($S_{C_x}$) is the survival function of the time to event $T_{x,i}$ (time to censoring $C_{x,i}$) in treatment group $x=A,B$.
}

{\small{\textsc{Proof.}}}
This can easily be shown with Hellands proposition \cite{Andersen} and using the fact that $Y_{x}(u)/n_x \stackrel{{\cal{P}}}{\rightarrow} y_{x}(u) \equiv S_{T_x}(u) \cdot S_{C_x}(u)$.
\quad $\Box$
\ \\ \\
\textbf{Lemma 5.}
\textit{Let $X_{x,i}\coloneqq T_{x,i} \wedge C_{x,i}$. Then, for any $i \neq j$, the density $f_{X_{x,i}}(u)$ and survival function $S_{X_{x,i}}(u)$ of $X_{x,i}$ as well as the density $f_{X_{x,i} \wedge X_{x,j}}(u)$ of $X_{x,i} \wedge X_{x,j}$ are 
	\begin{align}\label{appA_eq10}
	\begin{split}
	f_{X_{x,i}}(u) &= f_{T_x}(u)S_{C_x}(u)+S_{T_x}(u)f_{C_x}(u) \\
	S_{X_{x,i}}(u) &= S_{T_x}(u)S_{C_x}(u)\\
	f_{X_{x,i} \wedge X_{x,j}}(u) &= 2 [f_{T_x}(u)S_{C_x}(u)+S_{T_x}(u)f_{C_x}(u)]S_{T_x}(u)S_{C_x}(u).
	\end{split}
	\tag{A.12}
	\end{align}
	where $f_{T_x}$ and $S_{T_x}$ ($f_{C_x}$ and $S_{C_x}$) are density and survival function of the time to event $T_{x,i}$ (time to censoring $C_{x,i}$) in treatment group $x=A,B$.
}

{\small{\textsc{Proof.}}}
Follows from elementary calculation with probability distributions using the independence of $T_{x,i}$, $T_{x,j}$, $C_{x,i}$ and $C_{x,j}$ for $i \neq j$.
\quad $\Box$
\ \\ \\
\textbf{Lemma 6.}
\textit{Let $\widehat{\Sigma}$ the process defined in (\ref{appA_eq01}). Then, pointwise in $s$, the limit $\Sigma(s) \coloneqq \lim_{n \to \infty} E[\widehat{\Sigma}(s)]$ exists. Under the contiguous alternatives $\Lambda_B(\cdot)= \omega_n \Lambda_A(\cdot)$ with $\omega_n=\exp(-n^{-1/2}\gamma)$ for some $\gamma \geq 0$, $\Sigma(s)$ amounts to 
	\begin{align}\label{appA_eq11}
	\begin{split}
	\Sigma^2(s) = \int_0^{\infty} F_{T_A}(s \wedge u) f_{C_A}(u) du + 2 \pi \cdot \int_0^{\infty} \sigma_A(s \wedge u) [f_{T_A}(u)S_{C_A}(u)+S_{T_A}(u)f_{C_A}(u)]S_{T_A}(u)S_{C_A}(u) du
	\end{split}
	\tag{A.13}
	\end{align}
	where $f_{T_x}$, $F_{T_x}$, $S_{T_x}$ ($f_{C_x}$, $F_{C_x}$, $S_{C_x}$) are density, distribution function and survival function of the time to event $T_{x,i}$ (time to censoring $C_{x,i}$) in treatment group $x=A,B$, where $\sigma_A(\cdot)$ is the function from (\ref{appA_eq09}), and $\pi = n_B/n_A$ the prefixed treatment arm allocation ratio.
}

{\small{\textsc{Proof.}}}
Since each of the families of random variables $\{N_{B,i}(s)\}_{i \in {\cal{N}}_B}$, $\{\widehat{\sigma}_x(s\wedge X_{x,i})\}_{i \in {\cal{N}}_B}$ and $\{\widehat{\sigma}_x(s \wedge X_{x,i} \wedge X_{x,j})\}_{i \neq j \in {\cal{N}}_B}$ is identically distributed, we have
\begin{align}\label{appA_eq12}
\begin{split}
E[\widehat{\Sigma}^2(s)] = E[N_{B,1}(s)] + \frac{n_B}{n_B n_A} E[\widehat{\sigma}_A(s\wedge X_{B,1})] + \frac{n_B(n_B-1)}{n_Bn_A}  E[\widehat{\sigma}_A(s\wedge X_{B,1} \wedge X_{B,2})].
\end{split}
\tag{A.14}
\end{align}
Thus, the limit $\Sigma(s) = \lim_{n \to \infty} E[\widehat{\Sigma}(s)]$ exists. 
Now additionally assume the contiguous alternatives. Then $\lim_{n \to \infty} E[N_{B,1}(s)] = E[N_{A,1}(s)]$, because $\lim_{n \to \infty} \Lambda_B = \Lambda_A$ under the contiguous alternatives. Moreover, by independence of the treatment groups and by (\ref{appA_eq09}), we conclude that
\begin{align}\label{appA_eq13}
\begin{split}
\lim_{n \to \infty} E[\widehat{\sigma}_A(s\wedge X_{B,i} \wedge X_{B,j})]
&= \lim_{n \to \infty} E_u[E[\widehat{\sigma}_A(s\wedge X_{B,i} \wedge X_{B,j})|X_{B,i} \wedge X_{B,j}=u]] \\
&= \lim_{n \to \infty} \int_0^{\infty} E[\widehat{\sigma}_A(s\wedge u)] f_{X_{B,i} \wedge X_{B,j}}(u) du \\
&= \int_0^{\infty} {\sigma}_A(s\wedge u) \lim_{n \to \infty} f_{X_{B,i} \wedge X_{B,j}}(u) du \\
&= \int_0^{\infty} {\sigma}_A(s\wedge u) f_{X_{A,i} \wedge X_{A,j}}(u) du.
\end{split}
\tag{A.15}
\end{align}
where the third equality holds by dominated convergence under application of the estimate from Lemma 1 in \cite{McKeague:1990} and the convergence of the expectation holds by Lemma 4.2 in \cite{Aalen:1976}. In particular, the second summand on the right hand side of (\ref{appA_eq12}) vanishes as $n \to \infty$. So we are done by supplying the explicit value of the density function $f_{X_{A,i} \wedge X_{A,j}}$ from (\ref{appA_eq10}). Notice that the last equality in (\ref{appA_eq13}) holds, because $\lim_{n \to \infty} \Lambda_B = \Lambda_A$ under the contiguous alternatives.
\quad $\Box$
\ \\ \\

\section*{Appendix B: Computation of Expectation of $\widehat{\Sigma}(\infty)$ }\label{appB}
In this section we derive the expectation of $\widehat{\Sigma}(\infty)$ as shown in Section \ref{sec:case_study}. Firstly, $E[\widehat{\Sigma}^2(\infty)]$ can be decomposed into
\begin{equation*}
n_B^{-1} E[N_B(s)] + n_B^{-1} n_A^{-1} \sum_{i,j \in {\cal{N}}_{B}} E[\widehat{\sigma}_A(\infty \wedge X_{B,i} \wedge X_{B,j})]
\end{equation*}
where the first summand is the same as the quantity in \eqref{eq:expectation_classical_variance}. For the second summand, we have
\begin{align*}
n_B^{-1} n_A^{-1} \sum_{i,j \in {\cal{N}}_{B}} E[\widehat{\sigma}_A(X_{B,i} \wedge X_{B,j})]
&=n_A^{-1} E[\widehat{\sigma}_A(X_{B,1})] + (n_B - 1) n_A^{-1} E[\widehat{\sigma}_A(X_{B,1} \wedge X_{B,2})]\\ 
&\rightarrow \pi \cdot E[\widehat{\sigma}_A(X_{B,1} \wedge X_{B,2})]
\end{align*}
as $n\to\infty$. The expectation on the right hand side is given by
\begin{align*}
&\int_0^{\infty} \widehat{\sigma}_A(u) dF_{X_{B,1} \wedge X_{B,2}}(u)\\
=&\int_0^{\infty} \widehat{\sigma}_A(u)(1-S_{T_B}(u)S_{C_B}(u)) d(S_{T_B}S_{C_B})(u) - \int_0^{\infty} \widehat{\sigma}_A(u)(1+S_{T_B}(u)S_{C_B}(u)) d(S_{T_B}S_{C_B})(u)\\
=&- 2\int_0^{\infty} \widehat{\sigma}_A(u)S_{T_B}^2(u)S_{C_B}(u) dS_{C_B}(u) - 2\int_0^{\infty} \widehat{\sigma}_A(u)S_{T_B}(u)S_{C_B}^2(u) dS_{T_B}(u)\\
=&2\int_0^{\infty} \widehat{\sigma}_A(u)S_{T_B}^2(u)S_{C_B}(u) dF_{C_B}(u) + 2\int_0^{\infty} \widehat{\sigma}_A(u)S_{T_B}(u)S_{C_B}^2(u) dF_{T_B}(u)
\end{align*}
where we applied the product rule several times.

\newpage

\section*{Appendix C: Empirical type I error rates in scenarios with high survival rates}\label{appC}

\begin{table}[!h]
	\caption{Empirical type I error rates ($\alpha_{new}$ and $\alpha_{LR}$) and powers ($1-\beta_{new}$ and $1-\beta_{LR}$) for the new test and for the classical two--sample log--rank test, respectively, under proportional hazards alternatives for Weibull distributed survival times with shape parameter $\kappa$ and 1--year survival rate $S_1=0.8$ in the control arm. Theoretical two--sided significance level: $5 \%$. Underlying total sample size $n$ (or $n'$) in Scenario 1 (or Scenario 2) calculated to achieve a theoretical power of $80 \%$  under the planning alternative $H_1:\Lambda_B = \omega_0 \cdot \Lambda_A$for the classical log--rank test using Schoenfeld's formula (or for the new test using formula (\ref{Sec_04_eq03})). \label{tab:LowEventRate}}
	
	\begin{center}
		\small
		\begin{tabular}{cc|ccccc|ccccc}
			\toprule
			& & \multicolumn{5}{c|}{Scenario 1} & \multicolumn{5}{c}{Scenario 2} \\[0.14cm]
			$\kappa$  & $\omega_0$ & $n$ & $\alpha_{new}$ & $\alpha_{LR}$ & $1-\beta_{new}$ & $1-\beta_{LR}$ & $n'$ & $\alpha_{new}$ & $\alpha_{LR}$ & $1-\beta_{new}$ & $1-\beta_{LR}$  \\
			\midrule
			0.1  &   0.50    &   372    &   0.048   &   0.049   &   0.785   &   0.784  &  294 & 0.048 & 0.050 & 0.677 & 0.675    \\
			0.1  &   0.67    &   968    &   0.049   &   0.049   &   0.799   &   0.797  &  845 & 0.049 & 0.049 & 0.738 & 0.737    \\
			0.1  &   0.80    &  2736    &   0.049   &   0.049   &   0.799   &   0.798  & 2554 & 0.050 & 0.050 & 0.771 & 0.770    \\[0.14cm]
			0.25 &   0.50    &   308    &   0.049   &   0.048   &   0.787   &   0.785  &  253 & 0.047 & 0.049 & 0.696 & 0.696    \\
			0.25 &   0.67    &   766    &   0.047   &   0.049   &   0.800   &   0.798  &  693 & 0.045 & 0.046 & 0.756 & 0.754    \\
			0.25 &   0.80    &  2032    &   0.049   &   0.050   &   0.803   &   0.800  & 1995 & 0.049 & 0.049 & 0.786 & 0.783    \\[0.14cm]
			0.5  &   0.50    &   232    &   0.050   &   0.052   &   0.796   &   0.792  &  200 & 0.047 & 0.048 & 0.724 & 0.721    \\
			0.5  &   0.67    &   554    &   0.048   &   0.048   &   0.806   &   0.800  &  523 & 0.047 & 0.048 & 0.775 & 0.770    \\
			0.5  &   0.80    &  1386    &   0.045   &   0.045   &   0.806   &   0.800  & 1376 & 0.045 & 0.046 & 0.800 & 0.795    \\[0.14cm]
			0.75 &   0.50    &   182    &   0.047   &   0.048   &   0.801   &   0.798  &  161 & 0.047 & 0.050 & 0.731 & 0.732    \\
			0.75 &   0.67    &   426    &   0.048   &   0.049   &   0.810   &   0.801  &  412 & 0.048 & 0.049 & 0.790 & 0.781    \\
			0.75 &   0.80    &  1048    &   0.048   &   0.048   &   0.814   &   0.803  & 1054 & 0.047 & 0.048 & 0.817 & 0.806    \\[0.14cm]
			1    &   0.50    &   146    &   0.048   &   0.051   &   0.805   &   0.800  &  141 & 0.046 & 0.050 & 0.784 & 0.782    \\
			1    &   0.67    &   344    &   0.050   &   0.050   &   0.814   &   0.802  &  354 & 0.049 & 0.048 & 0.827 & 0.814    \\
			1    &   0.80    &   984    &   0.053   &   0.052   &   0.818   &   0.801  &  892 & 0.051 & 0.050 & 0.837 & 0.819    \\[0.14cm]
			1.25 &   0.50    &   120    &   0.047   &   0.052   &   0.800   &   0.794  &  110 & 0.046 & 0.050 & 0.760 & 0.756    \\
			1.25 &   0.67    &   288    &   0.050   &   0.051   &   0.817   &   0.800  &  283 & 0.050 & 0.052 & 0.806 & 0.791    \\
			1.25 &   0.80    &   748    &   0.052   &   0.050   &   0.823   &   0.801  &  792 & 0.052 & 0.051 & 0.826 & 0.804    \\[0.14cm]
			1.5  &   0.50    &   102    &   0.047   &   0.051   &   0.806   &   0.799  &   94 & 0.046 & 0.053 & 0.766 & 0.759    \\
			1.5  &   0.67    &   252    &   0.049   &   0.050   &   0.818   &   0.798  &  246 & 0.051 & 0.051 & 0.809 & 0.790    \\
			1.5  &   0.80    &   688    &   0.051   &   0.051   &   0.817   &   0.799  &  691 & 0.050 & 0.050 & 0.814 & 0.799    \\[0.14cm]
			2    &   0.50    &    80    &   0.043   &   0.057   &   0.797   &   0.798  &   71 & 0.036 & 0.056 & 0.707 & 0.735    \\
			2    &   0.67    &   212    &   0.048   &   0.056   &   0.803   &   0.799  &  202 & 0.046 & 0.054 & 0.785 & 0.776    \\
			2    &   0.80    &   644    &   0.051   &   0.052   &   0.815   &   0.802  &  631 & 0.049 & 0.052 & 0.800 & 0.793    \\[0.14cm]
			5    &   0.50    &    66    &   0.026   &   0.059   &   0.402   &   0.779  &   66 & 0.026 & 0.059 & 0.402 & 0.779    \\
			5    &   0.67    &   196    &   0.041   &   0.055   &   0.742   &   0.792  &  196 & 0.041 & 0.055 & 0.742 & 0.792    \\
			5    &   0.80    &   632    &   0.053   &   0.052   &   0.811   &   0.799  &  631 & 0.050 & 0.052 & 0.805 & 0.798    \\[0.14cm]
			\bottomrule
		\end{tabular}
	\end{center}
\end{table}

\newpage

\section*{Appendix D: Empirical type I error rates in scenarios with low survival rates}\label{appD}

\begin{table}[!h]
	\caption{Empirical type I error rates ($\alpha_{new}$ and $\alpha_{LR}$) and powers ($1-\beta_{new}$ and $1-\beta_{LR}$) for the new test and for the classical two--sample log--rank test, respectively, under proportional hazards alternatives for Weibull distributed survival times with shape parameter $\kappa$ and 1--year survival rate $S_1=0.2$ in the control arm. Theoretical two--sided significance level: $5 \%$. Underlying total sample size $n$ (or $n'$) in Scenario 1 (or Scenario 2) calculated to achieve a theoretical power of $80 \%$  under the planning alternative $H_1:\Lambda_B = \omega_0 \cdot \Lambda_A$for the classical log--rank test using Schoenfeld's formula (or for the new test using formula (\ref{Sec_04_eq03})). \label{tab:HighEventRate}}
	
	\begin{center}
		\small
		\begin{tabular}{cc|ccccc|ccccc}
			\toprule
			& & \multicolumn{5}{c|}{Scenario 1} & \multicolumn{5}{c}{Scenario 2} \\[0.14cm]
			$\kappa$  & $\omega_0$ & $n$ & $\alpha_{new}$ & $\alpha_{LR}$ & $1-\beta_{new}$ & $1-\beta_{LR}$ & $n'$ & $\alpha_{new}$ & $\alpha_{LR}$ & $1-\beta_{new}$ & $1-\beta_{LR}$  \\
			\midrule
			0.1  &   0.50    &    92    &   0.051   &   0.053   &   0.815   &   0.804  &  79  & 0.047  &  0.055 & 0.736 & 0.736   \\
			0.1  &   0.67    &   252    &   0.049   &   0.051   &   0.814   &   0.799  & 235  & 0.048  &  0.052 & 0.782 & 0.770  \\
			0.1  &   0.80    &   768    &   0.051   &   0.051   &   0.817   &   0.804  & 743  & 0.050  &  0.048 & 0.802 & 0.790  \\[0.14cm]
			0.25 &   0.50    &    86    &   0.052   &   0.049   &   0.818   &   0.801  &  75  & 0.045  &  0.057 & 0.745 & 0.743  \\
			0.25 &   0.67    &   234    &   0.052   &   0.049   &   0.818   &   0.801  & 222  & 0.054  &  0.053 & 0.803 & 0.781  \\
			0.25 &   0.80    &   706    &   0.055   &   0.052   &   0.824   &   0.806  & 694  & 0.053  &  0.052 & 0.815 & 0.798  \\[0.14cm]
			0.5  &   0.50    &    78    &   0.040   &   0.058   &   0.818   &   0.808  &  71  & 0.033  &  0.057 & 0.748 & 0.761  \\
			0.5  &   0.67    &   214    &   0.055   &   0.054   &   0.833   &   0.805  & 207  & 0.053  &  0.055 & 0.821 & 0.788  \\
			0.5  &   0.80    &   654    &   0.053   &   0.051   &   0.832   &   0.803  & 650  & 0.055  &  0.051 & 0.830 & 0.800  \\[0.14cm]
			0.75 &   0.50    &    72    &   0.030   &   0.059   &   0.750   &   0.799  &  68  & 0.029  &  0.059 & 0.709 & 0.776  \\
			0.75 &   0.67    &   204    &   0.051   &   0.056   &   0.823   &   0.801  & 200  & 0.048  &  0.055 & 0.816 & 0.795  \\
			0.75 &   0.80    &   636    &   0.052   &   0.051   &   0.822   &   0.799  & 635  & 0.052  &  0.053 & 0.817 & 0.798  \\[0.14cm]
			1    &   0.50    &    68    &   0.027   &   0.059   &   0.599   &   0.789  &  66  & 0.026  &  0.060 & 0.568 & 0.775  \\
			1    &   0.67    &   198    &   0.044   &   0.055   &   0.784   &   0.797  & 197  & 0.039  &  0.054 & 0.764 & 0.792  \\
			1    &   0.80    &   632    &   0.053   &   0.052   &   0.815   &   0.798  & 632  & 0.053  &  0.052 & 0.815 & 0.798  \\[0.14cm]
			1.25 &   0.50    &    68    &   0.027   &   0.058   &   0.506   &   0.793  &  66  & 0.026  &  0.059 & 0.476 & 0.779  \\
			1.25 &   0.67    &   198    &   0.042   &   0.056   &   0.760   &   0.798  & 196  & 0.041  &  0.055 & 0.751 & 0.792  \\
			1.25 &   0.80    &   632    &   0.053   &   0.052   &   0.812   &   0.799  & 631  & 0.050  &  0.052 & 0.808 & 0.798  \\[0.14cm]
			1.5  &   0.50    &    66    &   0.026   &   0.059   &   0.424   &   0.779  &  66  & 0.026  &  0.059 & 0.424 & 0.779  \\
			1.5  &   0.67    &   196    &   0.041   &   0.055   &   0.744   &   0.792  & 196  & 0.041  &  0.055 & 0.744 & 0.792  \\
			1.5  &   0.80    &   632    &   0.053   &   0.052   &   0.811   &   0.799  & 631  & 0.050  &  0.052 & 0.805 & 0.798  \\[0.14cm]
			2    &   0.50    &    66    &   0.026   &   0.059   &   0.403   &   0.779  &  66  & 0.026  &  0.059 & 0.403 & 0.779  \\
			2    &   0.67    &   196    &   0.041   &   0.055   &   0.742   &   0.792  & 196  & 0.041  &  0.055 & 0.742 & 0.792  \\
			2    &   0.80    &   632    &   0.053   &   0.052   &   0.811   &   0.799  & 630  & 0.052  &  0.052 & 0.809 & 0.798 \\[0.14cm]
			5    &   0.50    &    66    &   0.026   &   0.059   &   0.402   &   0.779  &  66  & 0.026  &  0.059 & 0.402 & 0.779  \\
			5    &   0.67    &   196    &   0.041   &   0.055   &   0.742   &   0.792  & 195  & 0.037  &  0.055 & 0.724 & 0.788  \\
			5    &   0.80    &   632    &   0.053   &   0.052   &   0.811   &   0.799  & 631  & 0.050  &  0.052 & 0.805 & 0.798  \\[0.14cm]
			\bottomrule
		\end{tabular}
	\end{center}
\end{table}

\section{Supporting information}
% Include only the SI item label in the paragraph heading. Use the \nameref{label} command to cite SI items in the text.
\paragraph{S1 File.}
\label{S1_File}
{\bf R code.}  Supplementary R code for the calculation of the test statistic $Z$ from \eqref{Sec_03_02_eq02} and corresponding two-sided $p$-values as well as functions for the analytic calculation of power and sample size (see \eqref{Sec_04_eq02} and \eqref{Sec_04_eq03}). Additionally, we provide another function to compute empirical type I and II errors on which our simulation study is based.

% Either type in your references using
% \begin{thebibliography}{}
% \bibitem{}
% Text
% \end{thebibliography}
%
% or
%
% Compile your BiBTeX database using our plos2015.bst
% style file and paste the contents of your .bbl file
% here. See http://journals.plos.org/plosone/s/latex for 
% step-by-step instructions.
% 

\end{document}